\documentclass{WileyMSP-template}
\usepackage{subcaption}
\usepackage{lineno}
\usepackage{amsmath}
\usepackage{float}

\usepackage[dvipsnames]{xcolor}
\usepackage{subcaption,amsmath,amssymb}
\begin{document}

\pagestyle{fancy}

\title{Metapinhole: Planar Fourier Optics Without Lenses}

\maketitle


\author{Mahmoud A. A. Abouelatta*}
\author{Karim Achouri}



\begin{affiliations}
M. A. A. A., K. A.\\
Laboratory for Advanced Electromagnetics and Photonics, École Polytechnique Fédérale de Lausanne (EPFL), 1015 Lausanne,
Switzerland\\
Email Address: mahmoud.abouelatta@epfl.ch


\end{affiliations}



\begin{abstract}
The 4f lens system is a standard Fourier-optics building block for angular control in optical platforms such as spatial light modulators, imaging systems, and data storage devices. This work presents the first nanoscale implementation of an equivalent system, achieving a five-orders-of-magnitude footprint reduction. By engineering the angular scattering response of metagratings, sharp-edge spatial filtering is realized through the interplay of angle-dependent two-dimensional dipolar resonances, Rayleigh anomalies, and Kerker-like dipolar cancellation. The metagrating functions as a high-efficiency low-pass and a controllable high-pass angular filter in transmission. In addition, diffraction-controllable angular invariance across the entire spatial Fourier space enables tunable band-pass filtering in reflection. This lens-free approach provides a compact, alignment-insensitive solution for spatial filtering in electromagnetic regimes where conventional lenses or pinholes are impractical or costly—such as the terahertz or infrared ranges—and facilitates spatial filtering for movable beams without complex mechanical adjustments. It also enables unprecedented single-shot multiplexing of diverse spatial filtering functions at distinct central wavelengths, and extends spatial filtering to signals with extremely low coherence lengths.
\end{abstract}


\section{Introduction}
Metasurfaces have enabled precise control over multiple degrees of freedom of light—frequency response, amplitude, phase profiles, and polarization~\cite{planar,metarev,synthesis}—by engineering the electromagnetic behavior of subwavelength structures. Despite these advances, approaches for tailoring the entire angular response of light remain limited. Achieving comprehensive angular control at the nanoscale could transform spatial optical computing, opening the door to light-speed signal processing~\cite{Fourier_karim}. Significant research has investigated angular responses for specific functions such as retro-reflection~\cite{retroreflector}, spatial differentiation~\cite{differentiation}, and refraction control~\cite{refraction}, among others~\cite{review_analog,review_romain}.

A key challenge in this context lies in the complexity of diffractive metasurfaces—such as phase gradient metasurfaces—which typically require numerous elements per unit cell, complicating fabrication. Metagratings address this limitation by reducing the unit cell to only one or a few elements, greatly simplifying fabrication while allowing rigorous analysis through diffraction theory and Floquet expansion for multiple ports. This structural simplicity is coupled with expanded design freedom for tailoring angular light responses, enabling advanced functionalities such as parallel computing~\cite{fleury}, extreme beam steering~\cite{metagrating_alu}, solving integral equations~\cite{integral}, pulse shaping~\cite{spatiotemp,spatiotemporal_exp,spatiotemp2}, perfect coherent diffraction~\cite{coherent}, holograms~\cite{hologram}, asymmetric angular response~\cite{step}, and Stokes parameters determination~\cite{stokes}.

Despite significant advances in metasurfaces and metagratings that replace bulky optical components with compact nanoscale systems, a planar equivalent to the standard 4f Fourier optics setup remains unavailable. There is an urgent need for such a system—not only to achieve significant miniaturization but also to eliminate dependence on costly or impractical lenses and pinholes, especially in electromagnetic regimes like the infrared and terahertz. Furthermore, it would enable effective spatial filtering of light signals with much shorter coherence lengths. A planar 4f equivalent would also enable position-invariant optical filtering for applications involving movable beams and allow unprecedented multiplexing of multiple filtering functions across different wavelength channels.
While such a nanoscale system is essential for advancing many photonics applications—including imaging, optical noise filtering, spatial light modulation, light display, and optical data storage—this study is, to the best of our knowledge, the first to explore this critical gap.

\begin{figure}[h]
\centering
\includegraphics[width=0.9\textwidth]{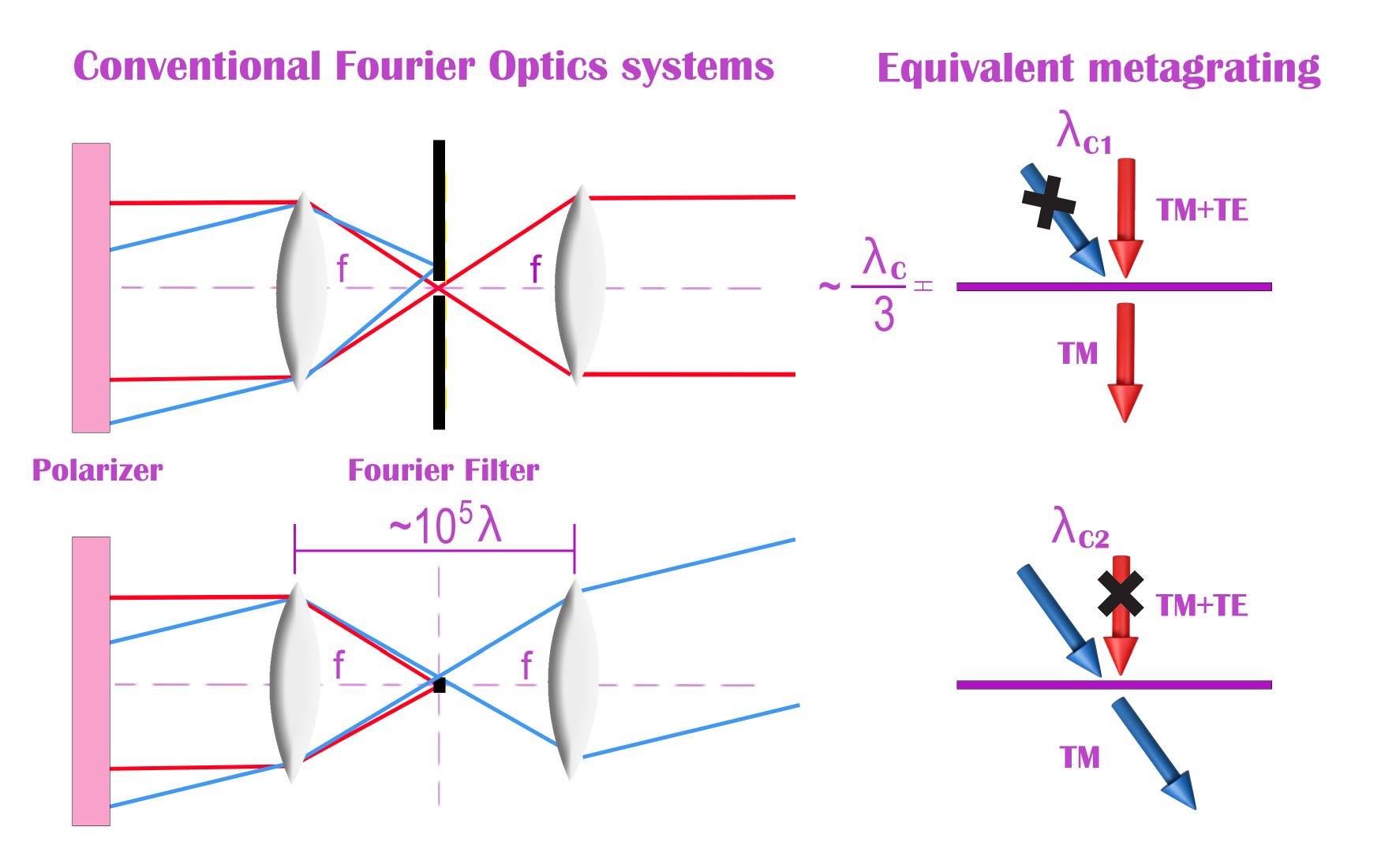}
\caption{Visual illustration of the equivalence between a standard 4f system and the proposed metagrating, achieving five orders of magnitude miniaturization. Moreover, it eliminates the need for pinholes or lenses, which are often impractical or costly in electromagnetic regimes such as terahertz, X-ray, or infrared, and is insensitive to alignment or position (i.e., the incident beam can be shifted without affecting the metagrating’s function). Additionally, it enables unprecedented multiplexing of different filtering functions, such as low-pass and high-pass, at distinct central wavelengths ($\lambda_{\rm C1}$ and $\lambda_{\rm C2}$), which is not feasible in a conventional 4f system.}
\label{fig:lens}
\end{figure}

As illustrated in Fig.~\ref{fig:lens}, the proposed metagrating can be considered equivalent to two conventional 4f Fourier optics systems—each consisting of 4f lens arrangements and Fourier filters for low-pass and high-pass angular filtering, respectively—while simultaneously enabling multiplexing of both responses at two distinct central wavelengths ($\lambda_{\rm C1}$ and $\lambda_{\rm C2}$). Moreover, it achieves a five-orders-of-magnitude reduction in device footprint while providing complete filtering of transverse electric (TE) polarization.

\section{Results}
\begin{figure}[h]
\centering
\includegraphics[width=0.8\textwidth]{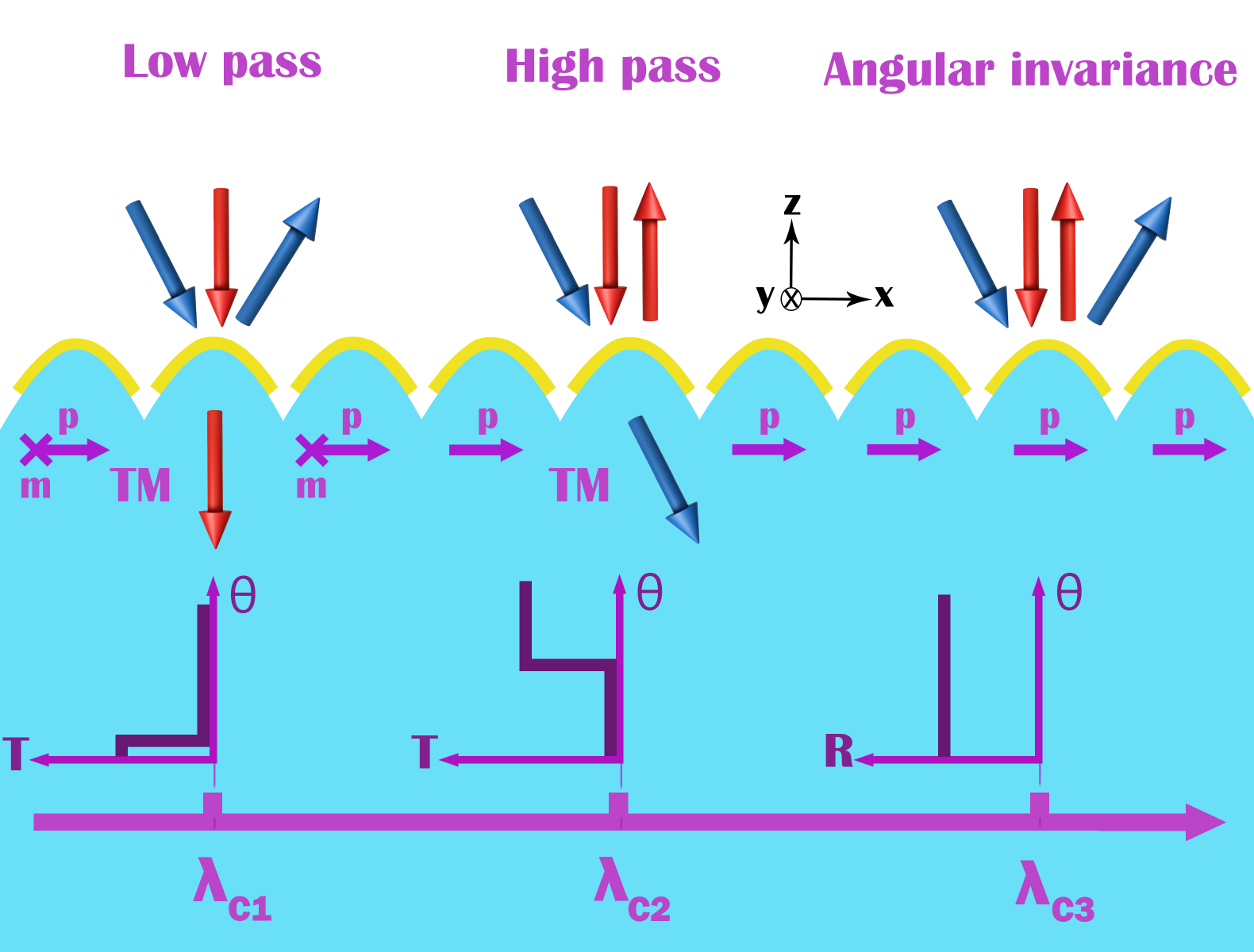}
\caption{Schematic of the proposed metagrating, composed of hat-like gold structures on silica with a periodic, inverted parabolic pattern. The functional objectives of three case studies—low-pass, high-pass angular filtering, and angular-invariant response—are illustrated at three distinct central wavelengths. At long wavelengths (e.g., $\lambda_{\rm C2}$ and $\lambda_{\rm C3}$), the metagrating behaves as an array of two-dimensional electric dipoles, while at shorter wavelengths (e.g., $\lambda_{\rm C1}$), it is analogous to an array of both two-dimensional electric and magnetic dipoles. The metagrating features a period of 680 nm, with a reversed parabolic height of 288 nm. The gold layer has a thickness of 78 nm, while the gap between the two hat-shaped elements measures 80 nm.}
\label{fig:meta}
\end{figure}

The proposed metagrating, depicted in Fig.~\ref{fig:meta}, consists of hat-like gold elements atop a silica layer with periodically reversed parabolic features. The structure has a period of 680 nm and, at three distinct wavelengths, it performs three distinct angular operations: low-pass, high-pass, and angular-invariant responses. Three-dimensional arrows indicate the response to normal and oblique plane-wave incidence, while the small arrows labeled \textbf{p} and \textbf{m} denote the equivalent two-dimensional electric and magnetic dipoles at each wavelength regime.

The metagrating topology exploits the interplay between diffraction and dipolar interactions to realize novel angular-filtering physics. Low-pass angular filtering is based on two-dimensional Kerker-like dipolar cancellation, which requires balanced electric and magnetic dipolar contributions within each unit cell. Metallic strips naturally emulate electric dipoles, and circular loops support magnetic dipoles; the hat-like elements introduced here hybridize these responses, providing a practical realization of this principle. 

High-pass angular filtering arises from the physics of diffractive electric dipoles. By precisely engineering the effective polarizability, transmission is strongly suppressed below the Rayleigh anomaly and sharply enhanced once diffraction sets in. This response can be obtained by intuition using arrays of curved metallic strips with subwavelength gaps. Below the diffraction threshold, the array behaves as an effective metallic surface that reflects incident waves. When diffraction begins, additional Floquet components appear, breaking the phase-matching condition causing transmission cancellation and leading to an abrupt rise in transmission. Hat-like elements exhibit this behavior at relatively long wavelengths, where the system acts as an array of two-dimensional electric dipoles, and at much longer wavelengths it functions as an effective metallic layer. A detailed formulation of the two-dimensional polarizability underlying this effect will be presented in future work.

Angular-invariant behavior occurs at the resonance of the two-dimensional electric dipolar mode. This occurs at a wavelength $\lambda_{\rm C3}$ larger than the grating period implying that only the 0th diffraction orders exist. Using a surface susceptibility tensor formalism for TM-polarized waves, the transmitted field can be expressed in terms of the tensor components, including the electric-electric susceptibility $\chi_{\rm ee}^{xx}$~\cite{invariance}. As shown in Eq.~\ref{eq:sus}, when the system approaches the dipolar resonance, $\chi_{\rm ee}^{xx}$ diverges, rendering the transmission effectively independent of the in-plane wavevector components as~\cite{invariance}
\begin{equation}
    \lim_{\chi_{\rm ee}^{xx}\to\infty} R(\theta) = \lim_{\chi_{\rm ee}^{xx}\to\infty}\dfrac{-(k_{z}^{2}\chi_{\rm ee}^{xx})}{k_{z}(k_{z}\chi_{\rm ee}^{xx}-2j)}=-1.
\label{eq:sus}
\end{equation}

Physically, this corresponds to a topology in which the collective scattering from each dipole cancels angle-dependent phase variations, producing wavevector-insensitive transmission. This mechanism provides a rigorous route to achieving angular invariance, as illustrated in Fig.~\ref{fig:meta}.

\begin{figure}[H]
\centering
\begin{subfigure}[b]{0.45\textwidth}
         \centering
         \includegraphics[width=\textwidth]{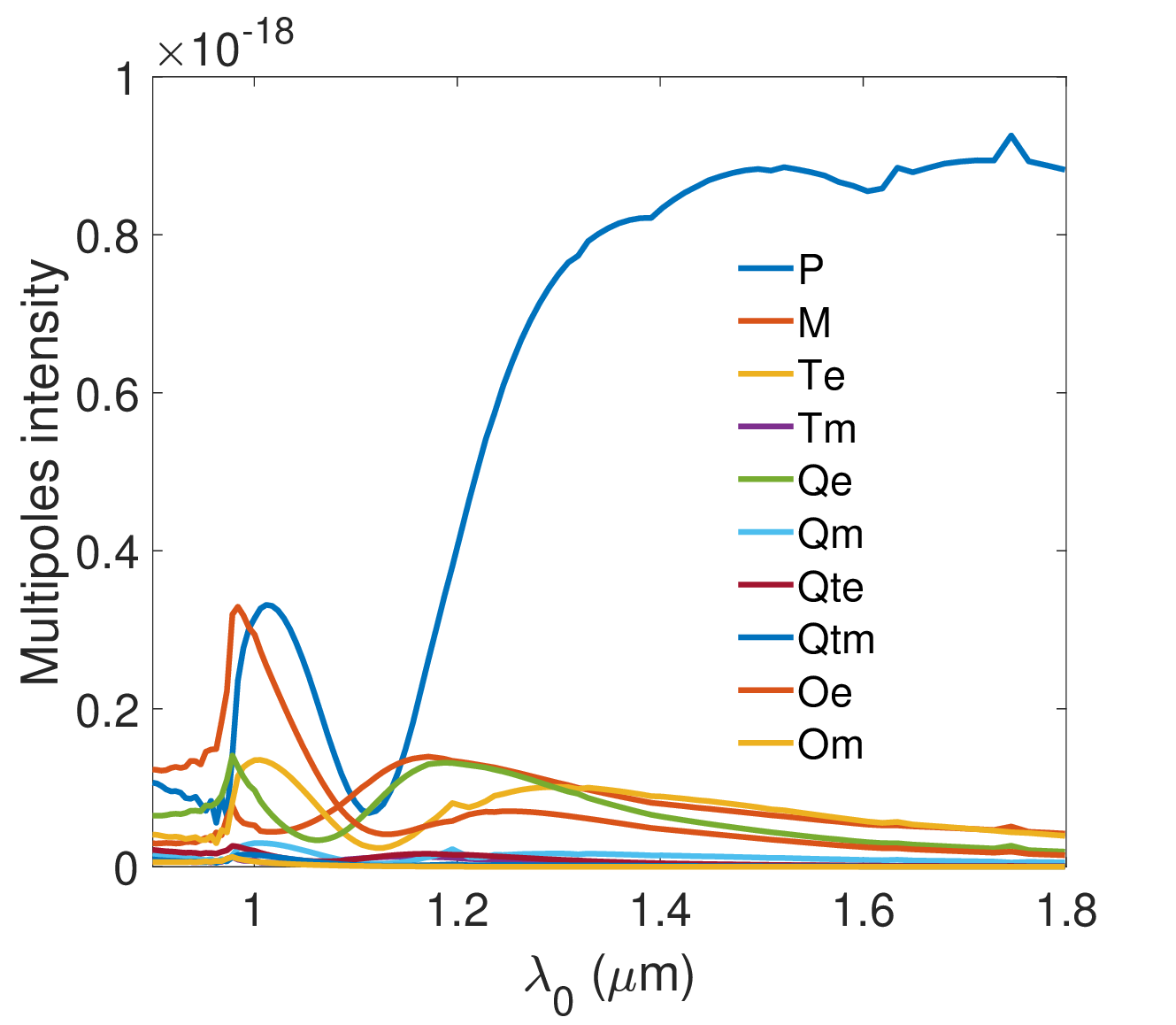}
         \caption{}
         \label{fig:N0_band_theo}
     \end{subfigure}
     \begin{subfigure}[b]{0.375\textwidth}
         \centering
         \includegraphics[width=\textwidth]{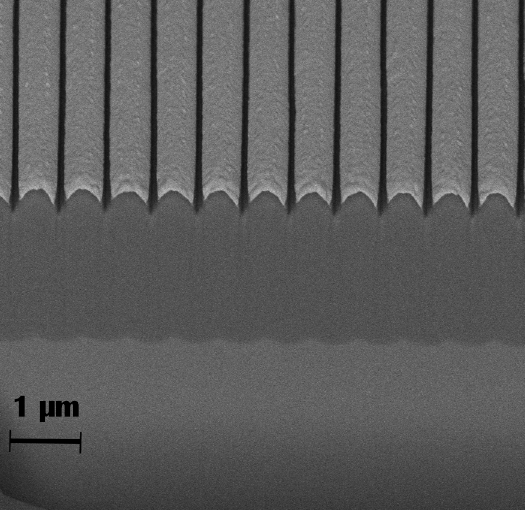}
         \caption{}
         \label{fig:N0_band_exp}
     \end{subfigure} 

\centering

     \caption{(a) Cartesian multipolar decomposition of the proposed metagrating, showing a Kerker-like condition at $\lambda_{0}$= 1000 nm between two-dimensional electric and magnetic dipoles, and a dominant electric dipole response at longer wavelengths. (b) Cross section of the 1 cm$^{2}$ fabricated sample, obtained by focused ion beam (FIB) milling and imaged using backscattered electrons in a scanning electron microscope (SEM). The bright regions on top correspond to the gold hat-like structures, while the dark and bright regions in the middle and bottom correspond to silica and silicon, respectively.}
\label{fig:FIB}
\end{figure}

A Cartesian multipolar decomposition of the introduced metagrating is shown in Fig.~\ref{fig:FIB}(a). Compared to its spherical counterpart, the Cartesian approach naturally incorporates periodic boundary conditions and, when one direction is invariant, efficiently models two-dimensional dipoles that are infinitely extended in that direction, such as cylindrical geometries. By contrast, while the spherical decomposition is exact, its application to periodic systems is cumbersome, and relating it to two-dimensional multipoles is far less straightforward.

As seen in Fig.~\ref{fig:FIB}(a), at approximately $\rm\lambda_{0} = 1\ \mu m$, the electric and magnetic dipoles are equal in magnitude and dominate the scattering response. This is also the point at which the first transmissive diffraction order emerges for incidence angles exceeding $2^{\circ}$. The simultaneous occurrence of Kerker-like dipolar balance and the onset of diffraction gives rise to a low-pass angular filtering effect, functioning as a “metapinhole.” At longer wavelengths—specifically above $\rm1.3\ \mu m$—the scattering is instead dominated by two-dimensional electric dipoles, consistent with the high-pass angular filtering regime discussed earlier.

The cross section of a fabricated 1 cm$^{2}$ sample is shown in Fig.~\ref{fig:FIB}(b). It was prepared via focused ion beam milling and imaged using backscattered electrons in a scanning electron microscope. In the resulting image, the bright array at the top corresponds to the hat-like structures, while the dark regions between them indicate nanoscale gaps, confirming they are well-separated. These gaps were created through two oblique gold evaporation steps, as described in the Supplemental Materials. Below this, the dark central region corresponds to silica, and the bright region at the bottom corresponds to silicon. The inverted parabolic array on silica was fabricated by first performing KOH etching of a silicon nitride mask on a silicon wafer, followed by wet thermal oxidation to convert the exposed surface into silica.

The fabricated sample is characterized using a supercontinuum source, polarizer, rotating holder, fiber coupler, optical fiber, and optical spectrum analyzer, as detailed in the Supplemental Materials. Figure~\ref{fig:dispersion} compares the simulated and measured angular dispersion of the 0$^{\rm th}$-order transmittance—the main output signal—for TM-polarized light shown in panels (a) and (b), and TE-polarized light shown in panels (c) and (d), demonstrating excellent agreement between simulations and measurements.

For TM polarization, the high-pass angular filtering regime at a wavelength of 1.4 $\rm \mu m$ achieves more than 30 dB suppression at small incidence angles compared to large angles. In contrast, the low-pass filtering regime at 1 $\rm \mu m$ exhibits over 10 dB reduction at large angles relative to small angles, as more clearly shown in Fig.~S1 of the supplementary materials, which presents the same data on a linear scale. At this wavelength and at normal incidence, the transmittance reaches –1 dB, demonstrating high efficiency despite the common assumption that metallic structures are inherently lossy and reflective—an efficiency attributed to Kerker-like dipolar cancellation at normal incidence. For TE polarization, the angular dispersion shows more than 30 dB suppression across the entire temporal and spatial spectrum, confirming a highly effective polarizing behavior.

\begin{figure}[H]
\centering
     \begin{subfigure}[b]{0.45\textwidth}
         \centering
         \includegraphics[width=\textwidth]{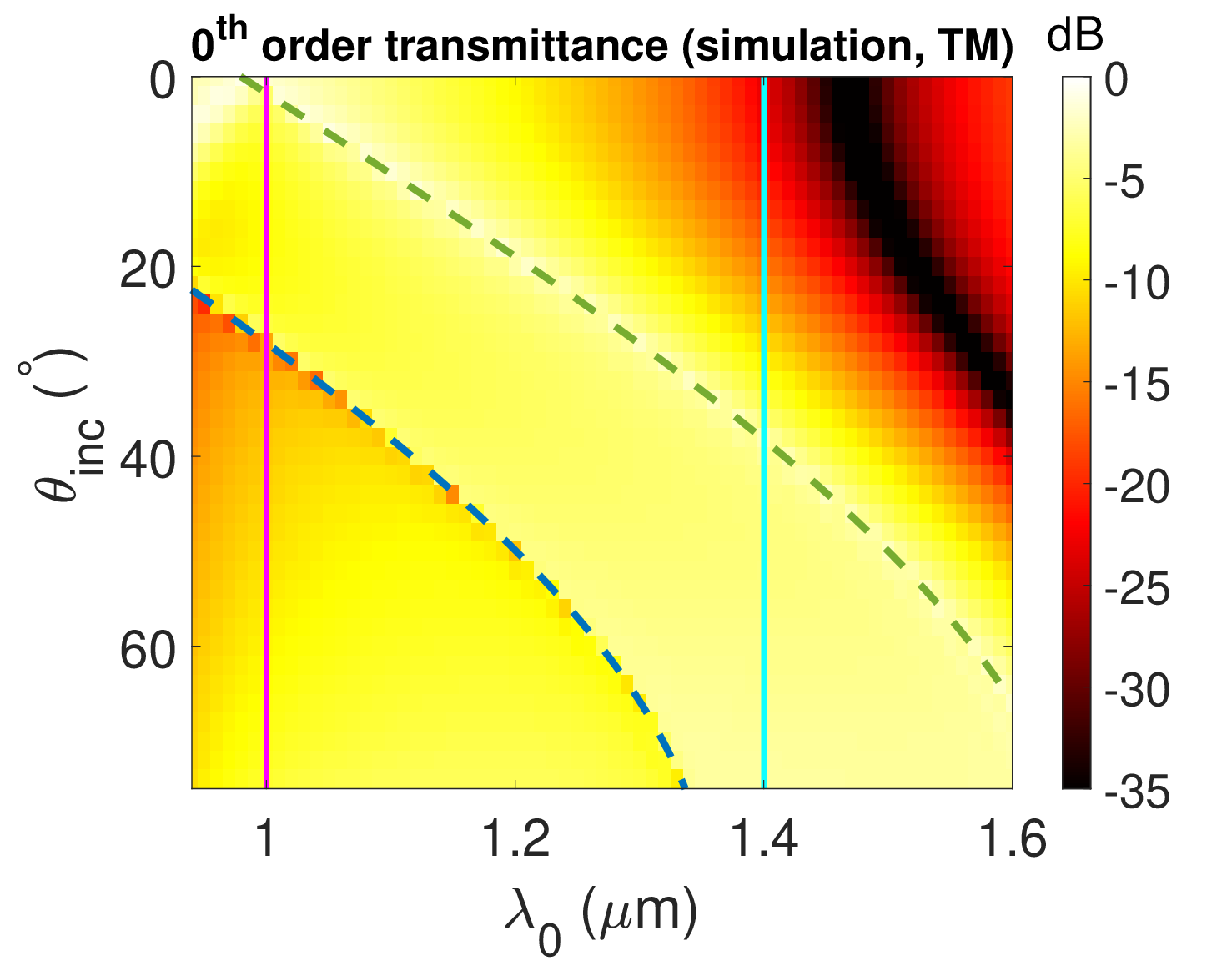}
         \caption{}
         \label{fig:N0_band_theo}
     \end{subfigure}
     \begin{subfigure}[b]{0.45\textwidth}
         \includegraphics[width=\textwidth]{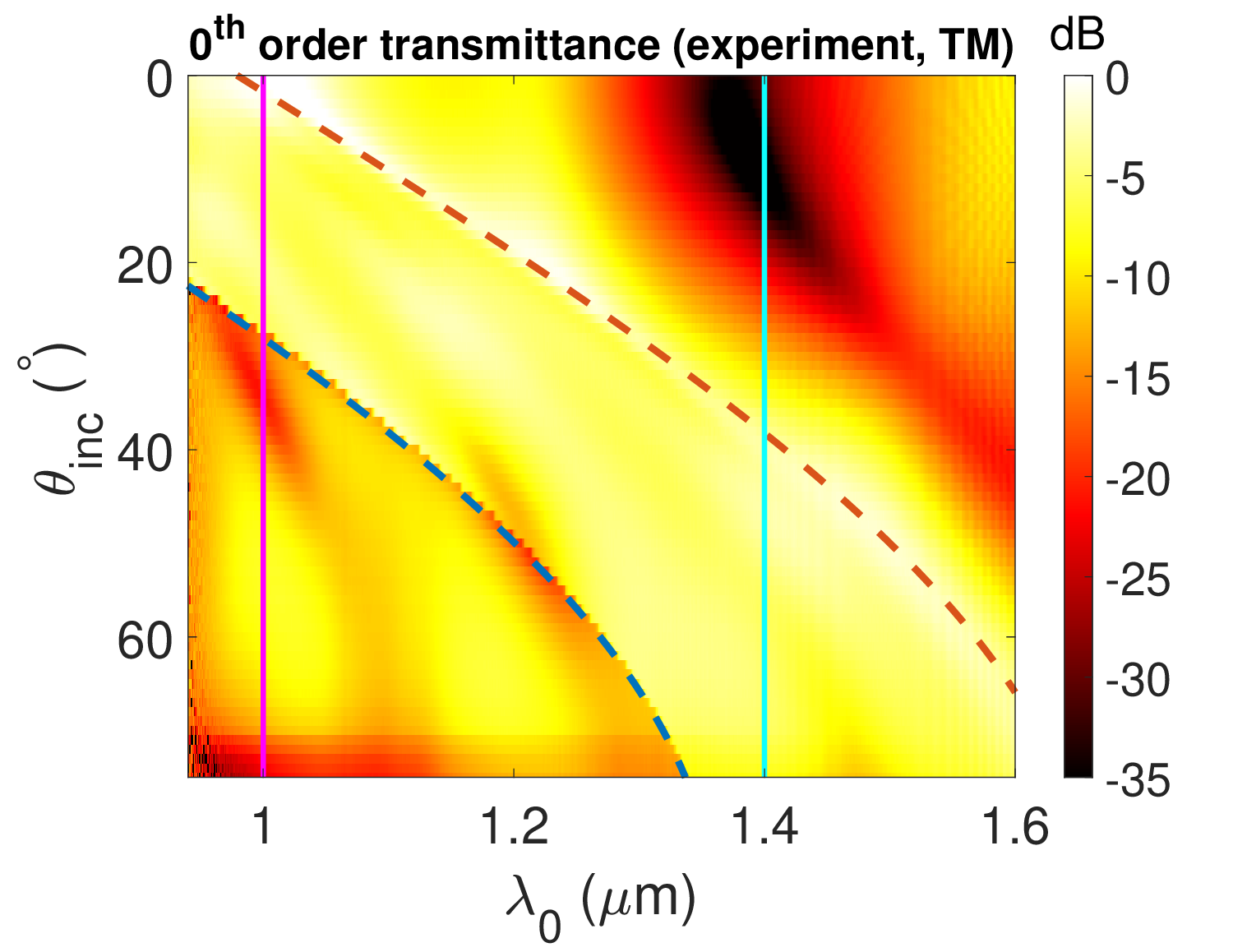}
         \caption{}
         \label{fig:N0_band_exp}
     \end{subfigure} 
     \begin{subfigure}[b]{0.45\textwidth}
         \centering
         \includegraphics[width=\textwidth]{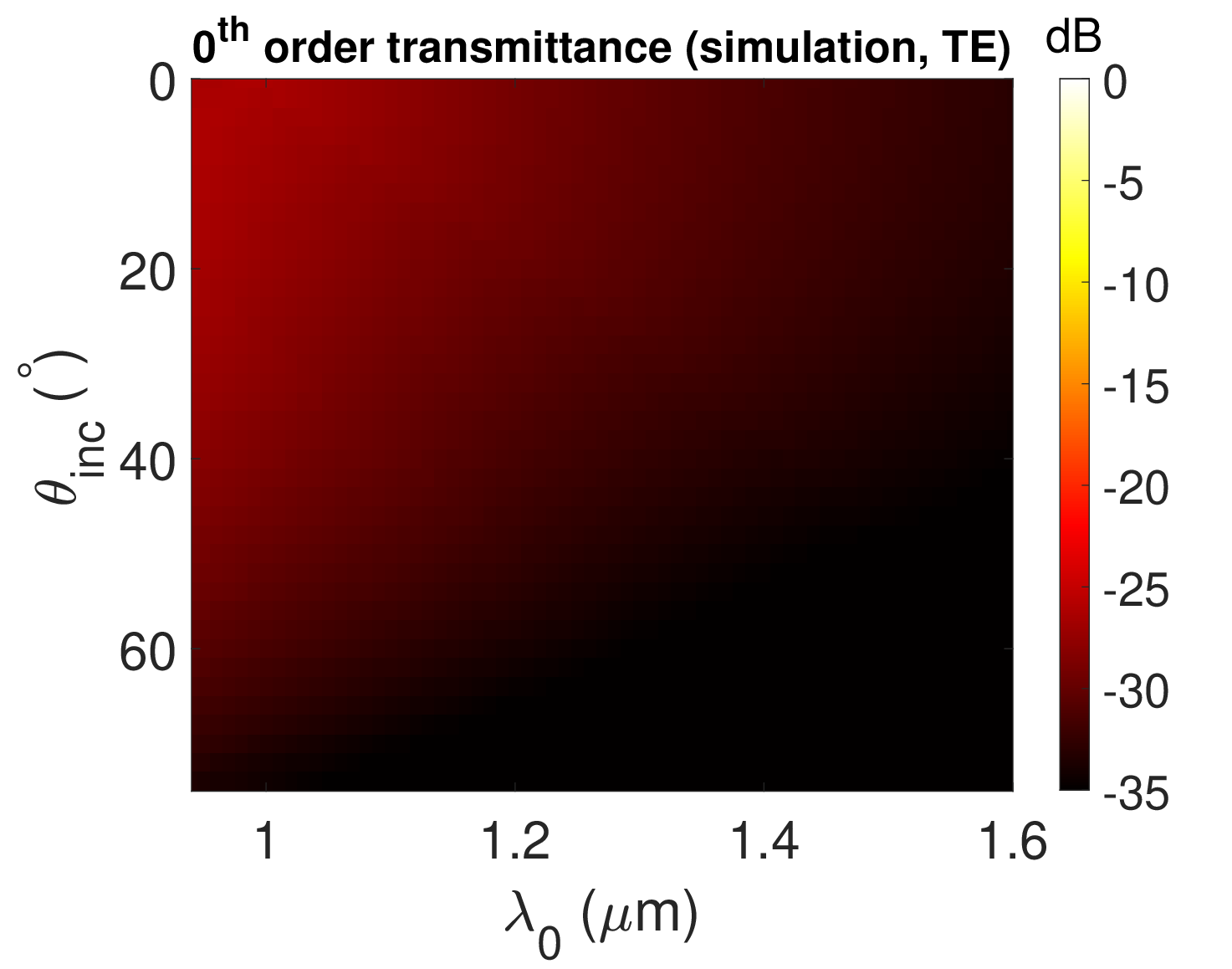}
         \caption{}
         \label{fig:N0_band_theo}
     \end{subfigure}
     \begin{subfigure}[b]{0.45\textwidth}
        \centering
         \includegraphics[width=\textwidth]{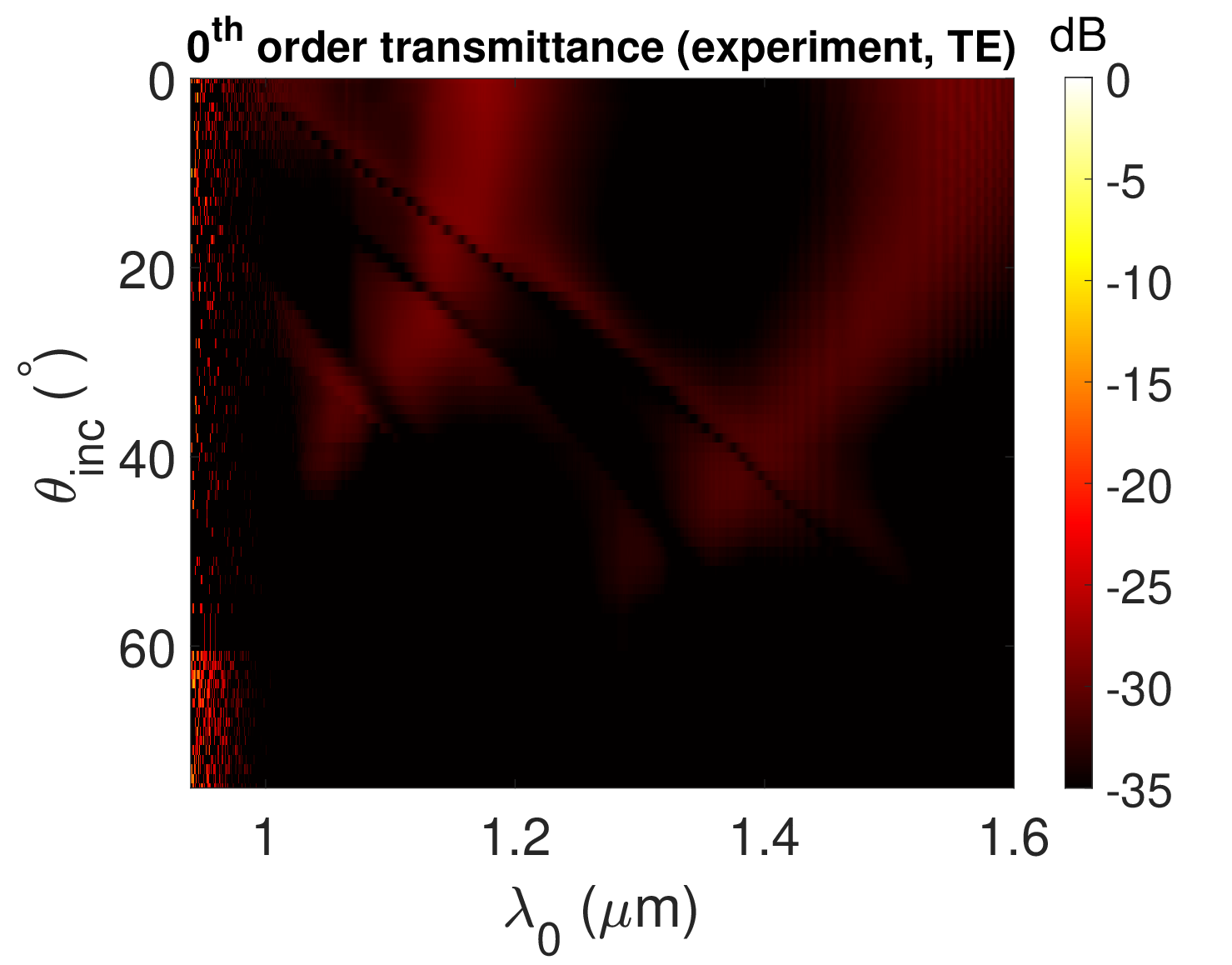}
         \caption{}
         \label{fig:N0_band_exp}
     \end{subfigure} 
\centering
     \caption{(a) Simulated and (b) measured angular dispersion of the $0^{\rm th}$ order transmittance for transverse magnetic (TM) polarization. (c) and (d) same as (a) and (b) for transverse electric (TE) polarization. The dashed lines correspond to the theoretical cutoff conditions for first-order diffraction in transmission and reflection, as given by the grating equation.}
\label{fig:dispersion}
\end{figure}

\begin{figure}[H]
\centering
\begin{subfigure}[b]{0.45\textwidth}
         \centering
         \includegraphics[width=\textwidth]{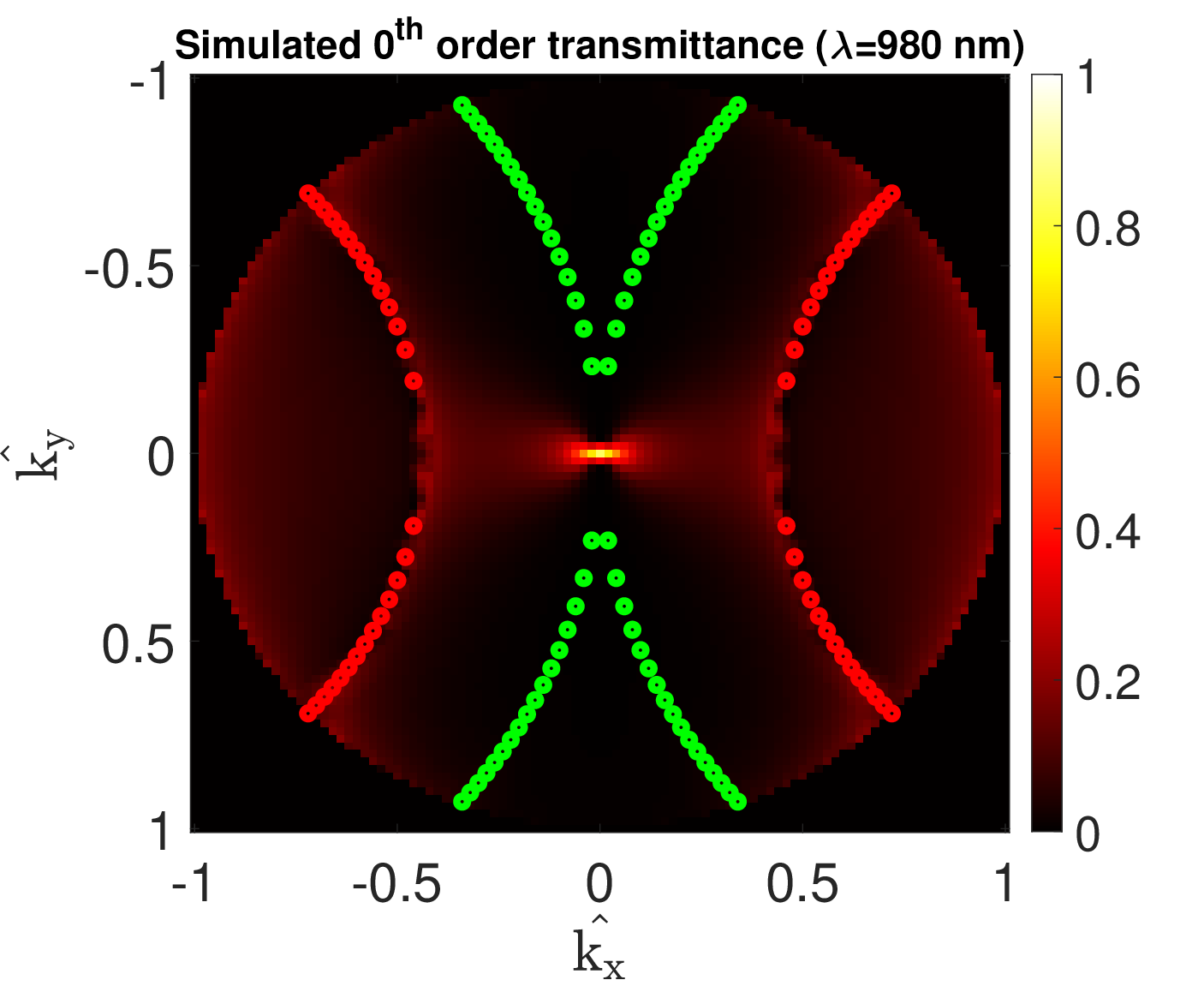}
         \caption{}
         \label{fig:N0_band_theo}
     \end{subfigure}
     \begin{subfigure}[b]{0.45\textwidth}
         \includegraphics[width=\textwidth]{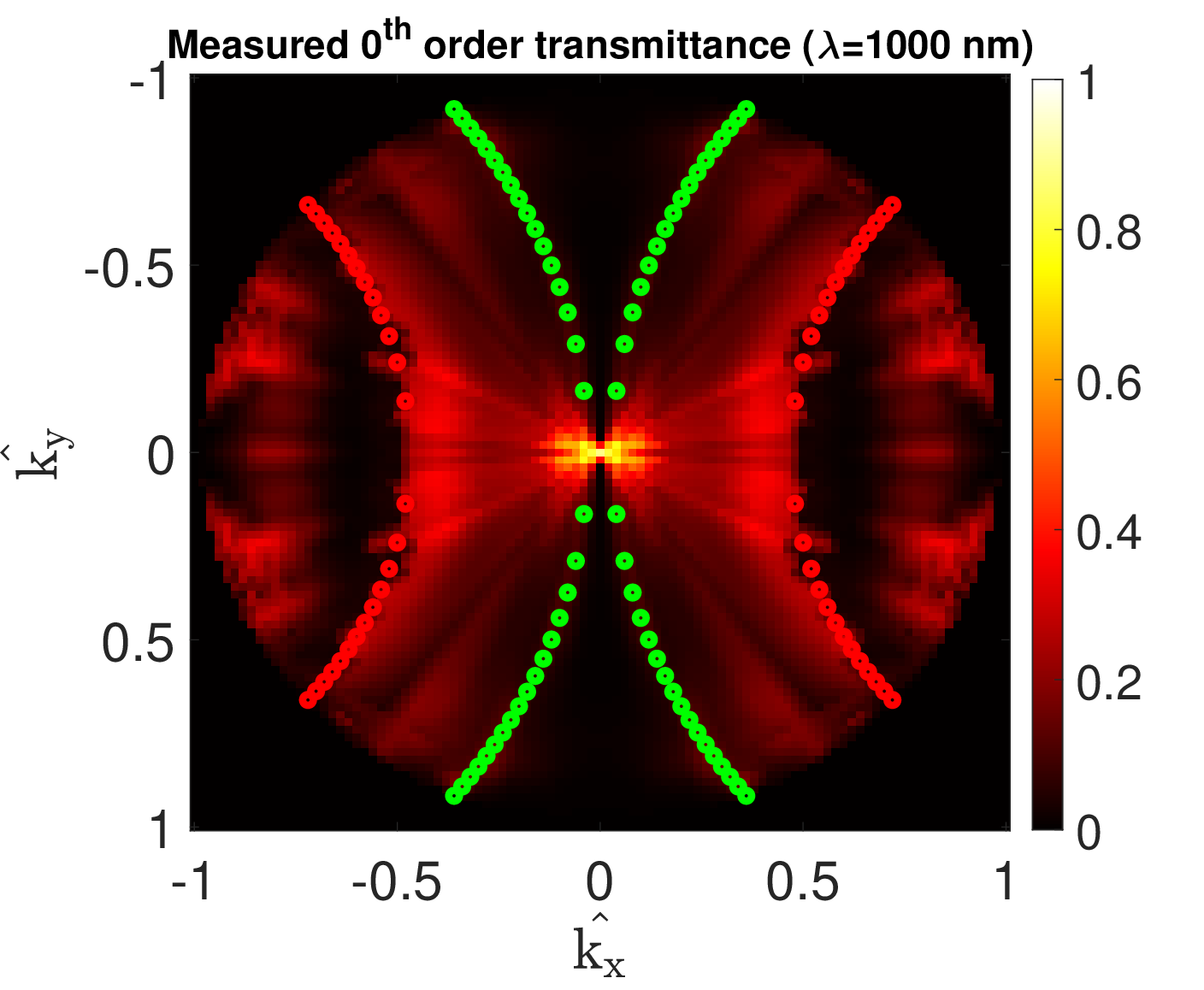}
         \caption{}
         \label{fig:N0_band_exp}
     \end{subfigure} 
     \begin{subfigure}[b]{0.45\textwidth}
         \centering
         \includegraphics[width=\textwidth]{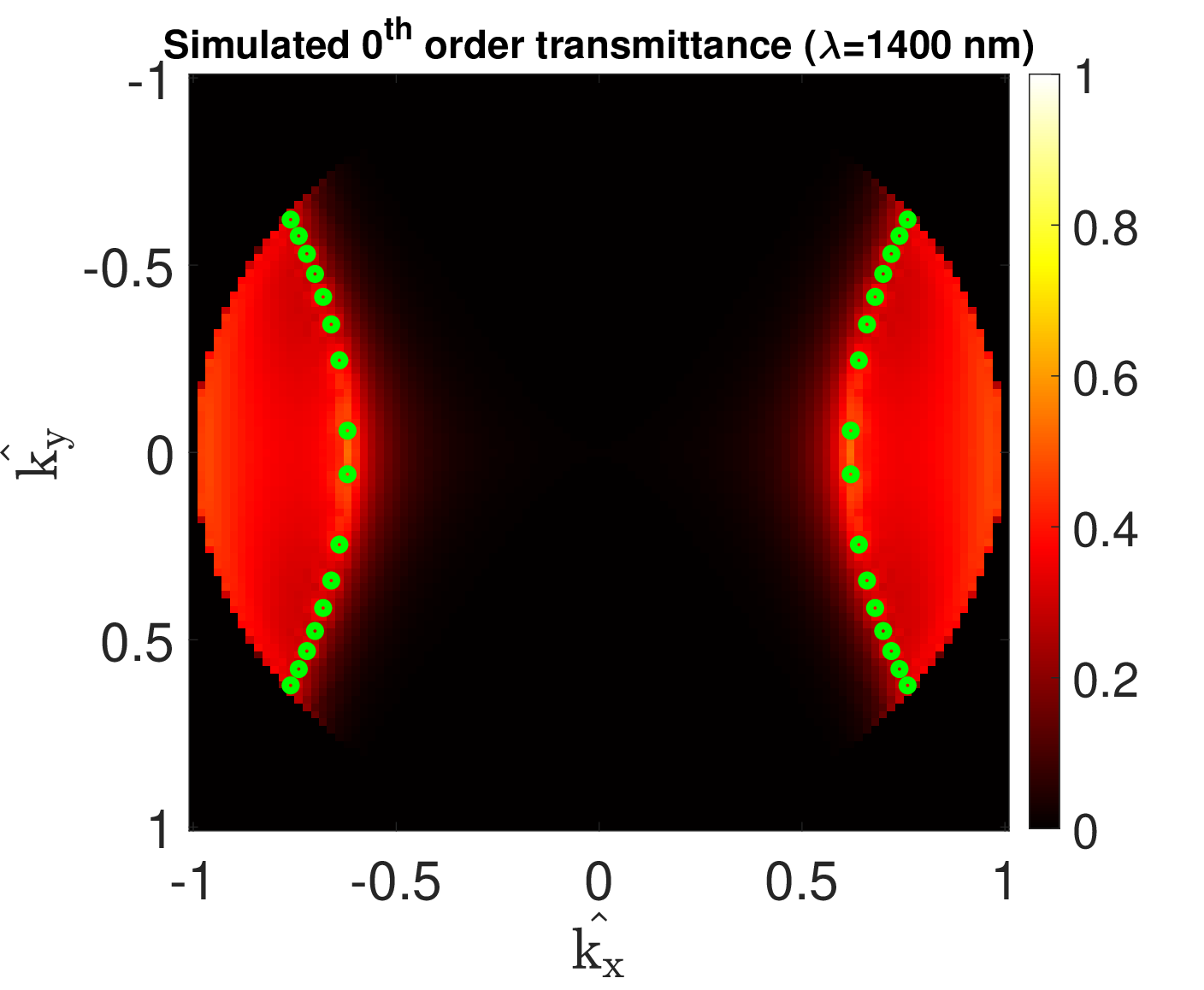}
         \caption{}
         \label{fig:N0_band_theo}
     \end{subfigure}
     \begin{subfigure}[b]{0.45\textwidth}
         \includegraphics[width=\textwidth]{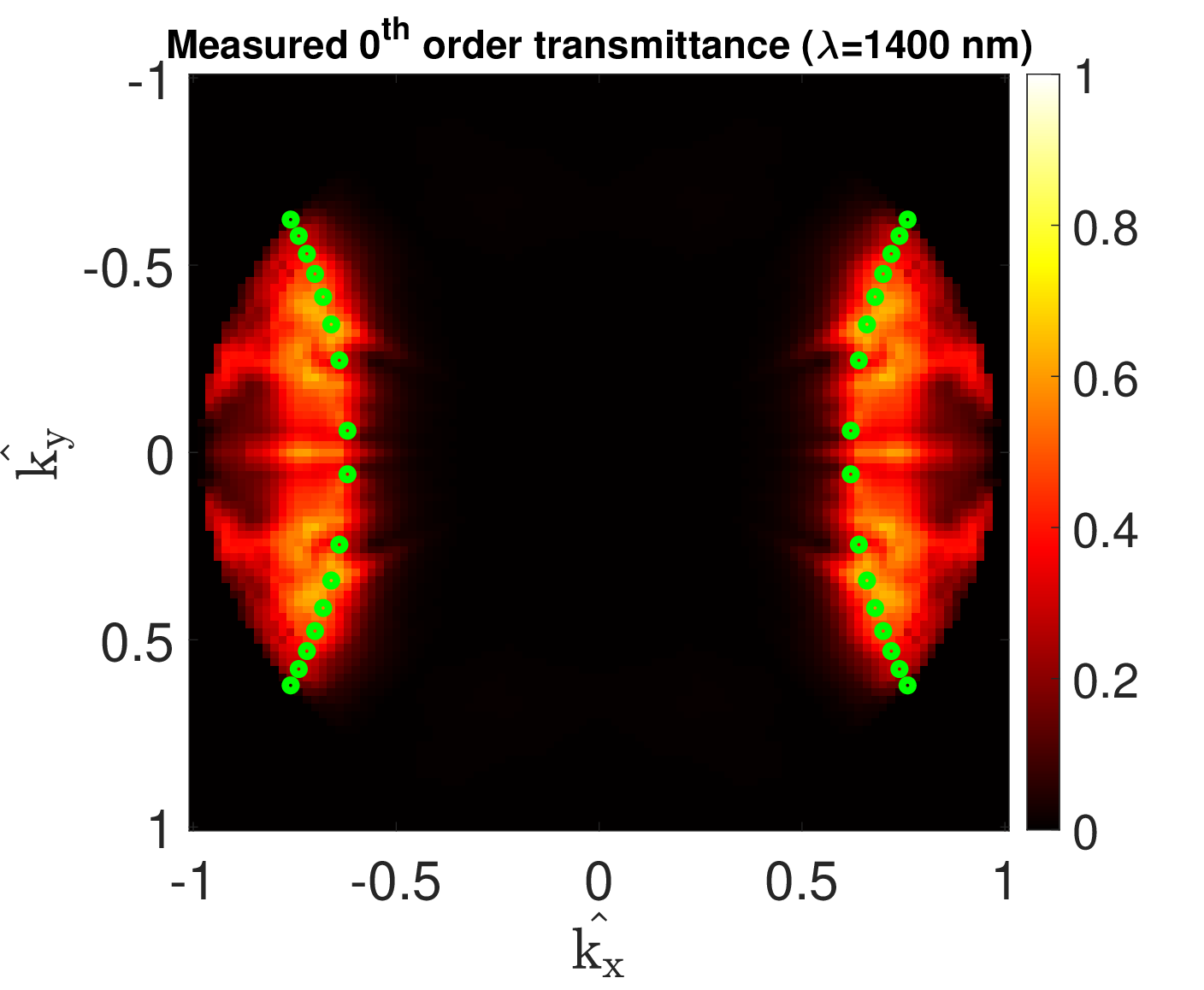}
         \caption{}
         \label{fig:N0_band_exp}
     \end{subfigure} 

\centering

     \caption{(a) Simulated and (b) measured $0^{\rm th}$-order transmittance for a low-pass angular filter in full spatial Fourier space (all three-dimensional angles) at $\lambda_{0}$ = 980 nm and 1000 nm, with a 20 nm shift between simulation and measurement. (c) Simulated and (d) measured $0^{\rm th}$-order transmittance for a high-pass angular filter in full spatial Fourier space at $\lambda_{0}$ = 1400 nm. The dashed lines indicate the theoretical cutoff boundaries for first-order diffraction in reflection and transmission, derived from conical diffraction theory~\cite{step}.}
\label{fig:spatial}
\end{figure}

Two lines at $\lambda_{0} = 1$ µm and 1.4 µm are highlighted in Fig.~\ref{fig:dispersion}, corresponding to low- and high-pass angular filtering effects, respectively. To characterize how the introduced metagrating operates across the full spatial Fourier space (i.e., as a Fourier filter) in both cases, Figs.~\ref{fig:spatial}(a)–(d) present the simulated and measured 0$^{\rm th}$-order transmittance at these wavelengths as a function of all possible propagating spatial frequencies along the $x$ and $y$ directions. The results—obtained by sweeping point-by-point across all three-dimensional spherical incident angles ($\theta$ and $\phi$), as described in the Supplemental Materials, and subsequently transforming into normalized spatial Fourier space ($\hat{k}_{x}$ vs. $\hat{k}_{y}$)—show strong agreement between simulation and experiment, confirming the practicality and broad potential of the metagrating for advanced planar Fourier-optics applications.

The low-pass spatial filtering effect is evident in Figs.~\ref{fig:spatial}(a) and (b), where the device functions effectively as a pinhole system. A small spectral shift of 20 nm between simulation and measurement is attributed to fabrication imperfections. In contrast, the high-pass filtering effect—controllable via the diffraction cutoff condition—is observed in Figs.~\ref{fig:spatial}(c) and (d), showing strong suppression of the 0$^{\rm th}$-order transmittance at small spatial frequencies, with dotted lines indicating the cutoff conditions for the first diffraction orders.

Residual measurement noise is likely due to the fact that, for incidence angles beyond the plane of symmetry of the metagrating, the TE/TM decomposition no longer provides a complete basis. In these cases, exact characterization of the response requires a full-rank $3\times 3$ tensor, since the mirror symmetry that underpins the decomposition is broken. In our simulations, however, the input is assumed to remain a pure TM mode as the azimuthal angle $\phi$ varies. This approximation does not strictly hold in measurements, leading to discrepancies. In addition, the results shown in Fig.~\ref{fig:spatial} are obtained from 1890 individually sampled measurements, which may have further contributed to the observed noise.

\begin{figure}[H]
\centering
\begin{subfigure}[b]{0.45\textwidth}
         \centering
         \includegraphics[width=\textwidth]{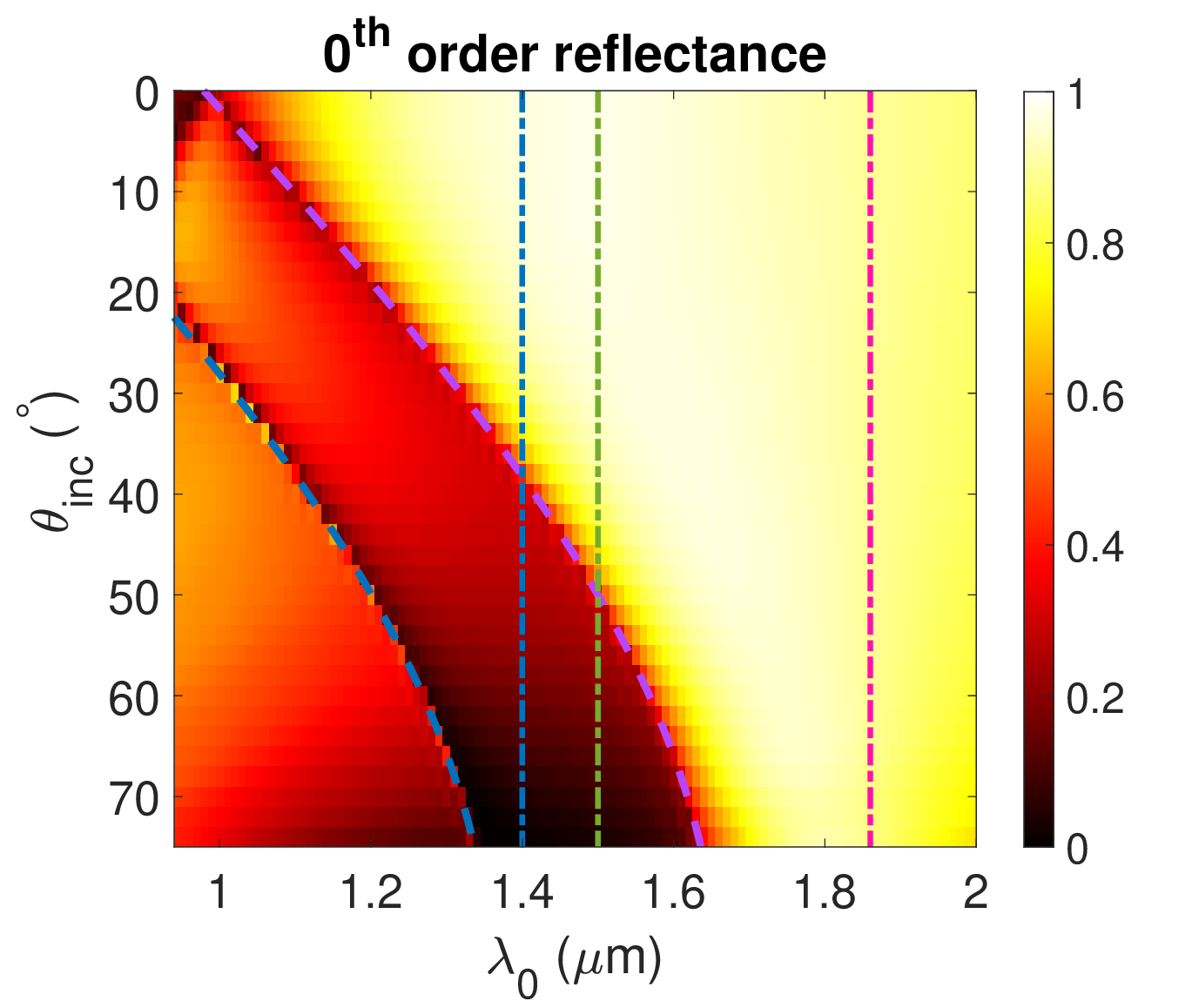}
         \caption{}
         \label{fig:N0_band_theo}
     \end{subfigure}
     \begin{subfigure}[b]{0.42\textwidth}
         -\includegraphics[width=\textwidth]{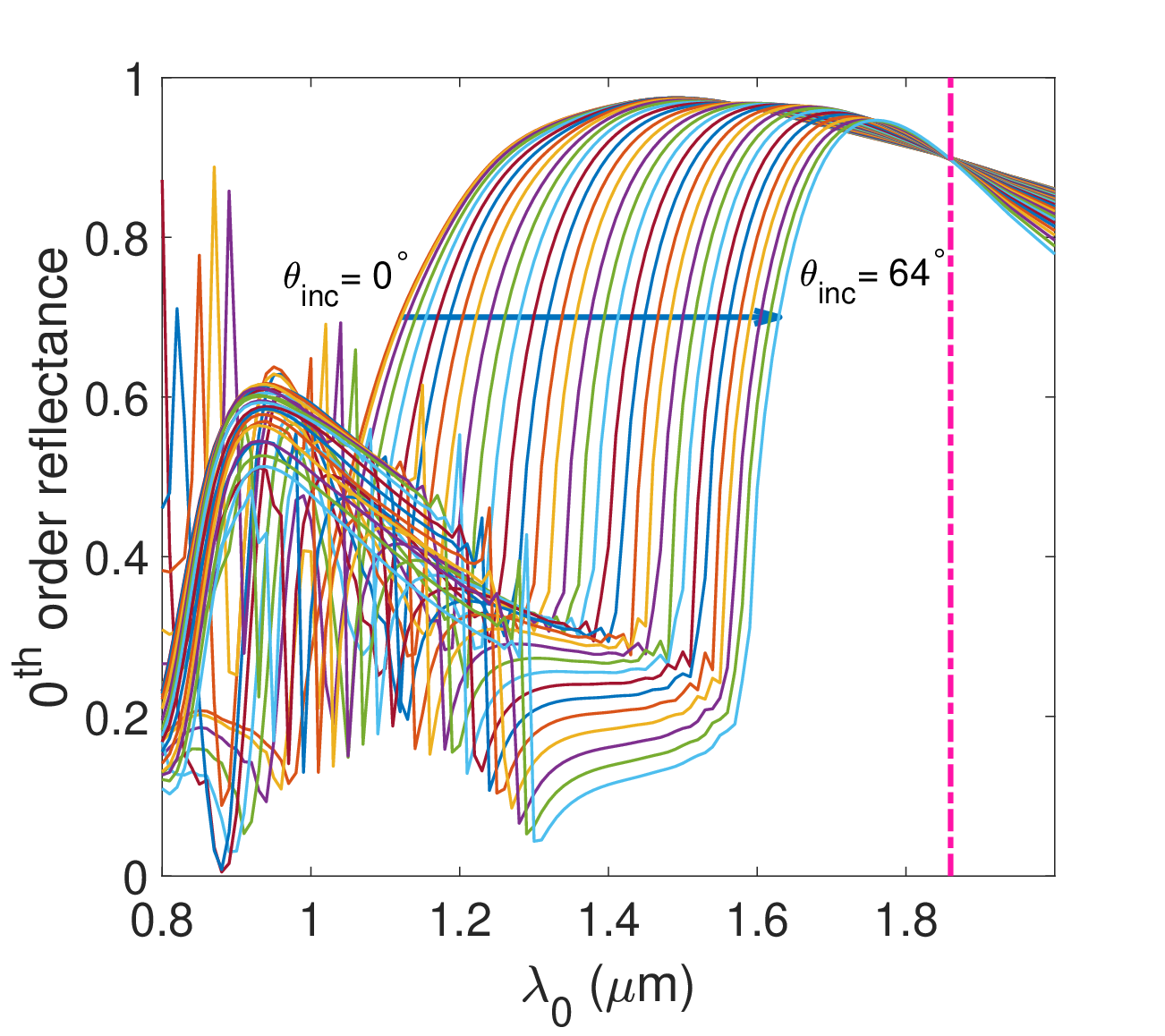}
         \caption{}
         \label{fig:N0_band_exp}
     \end{subfigure} 
     \begin{subfigure}[b]{0.3\textwidth}
         \centering
         \includegraphics[width=\textwidth]{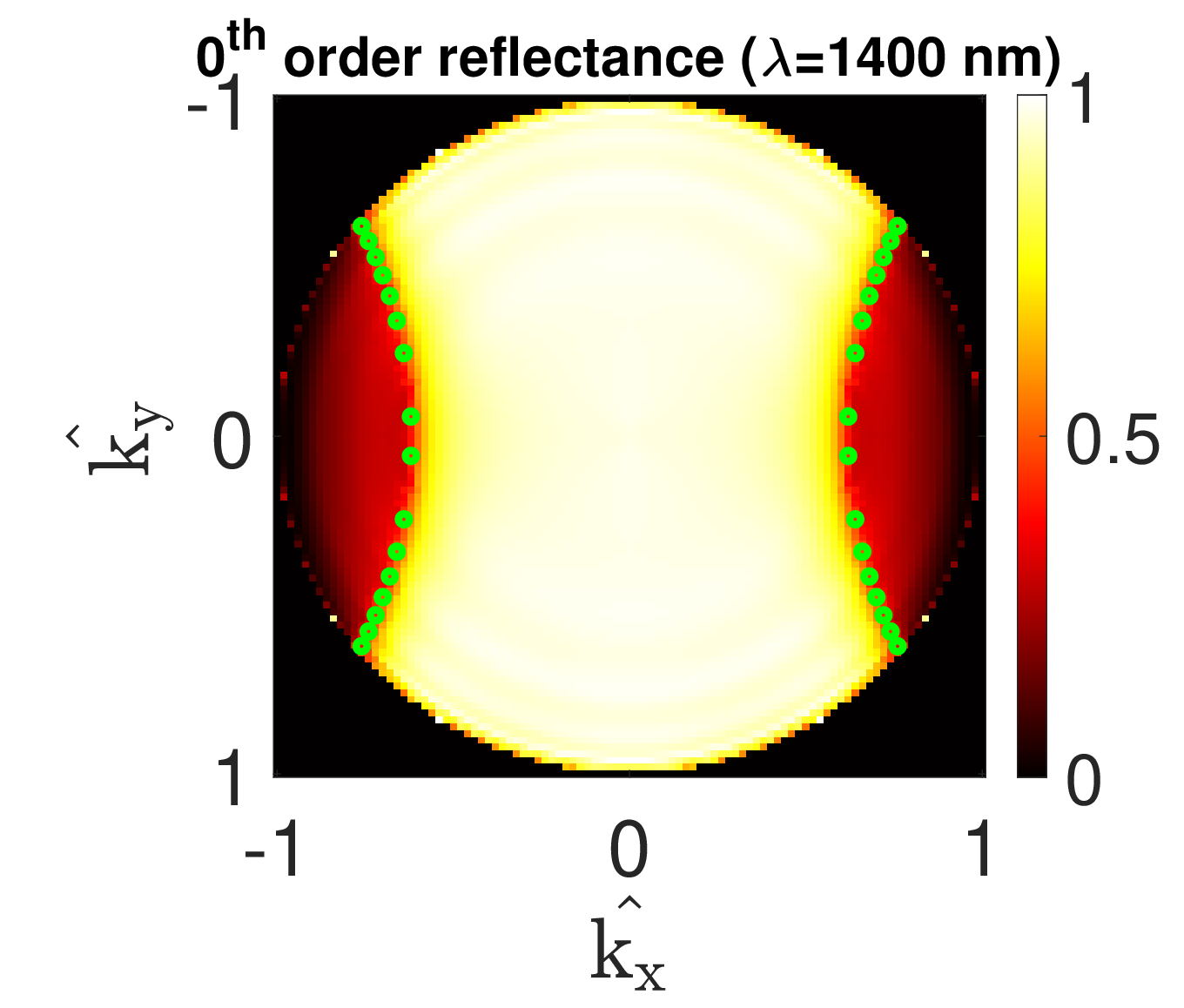}
         \caption{}
         \label{fig:N0_band_theo}
     \end{subfigure}
     \begin{subfigure}[b]{0.3\textwidth}
         \includegraphics[width=\textwidth]{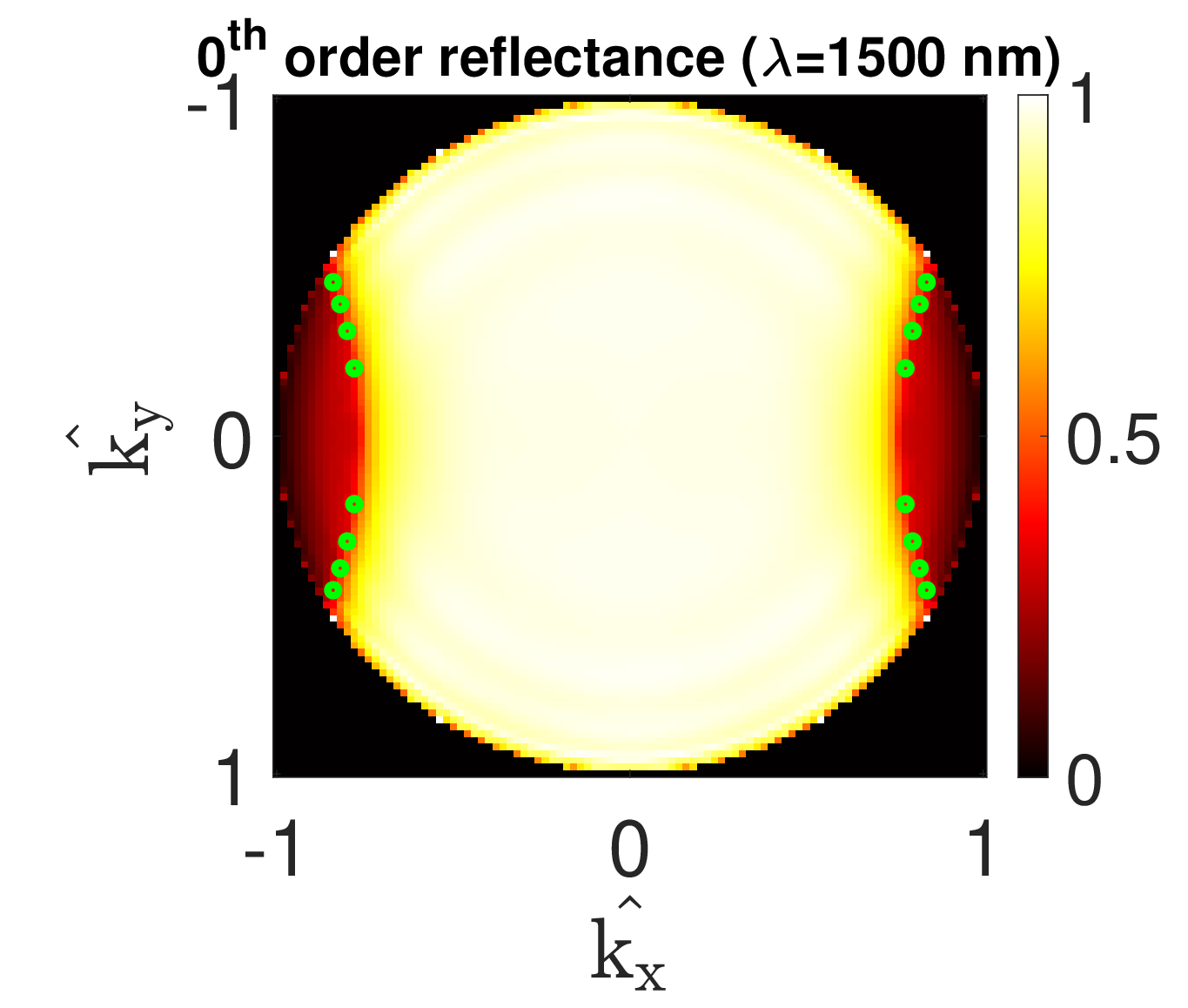}
         \caption{}
         \label{fig:N0_band_exp}
     \end{subfigure}
          \begin{subfigure}[b]{0.3\textwidth}
         \includegraphics[width=\textwidth]{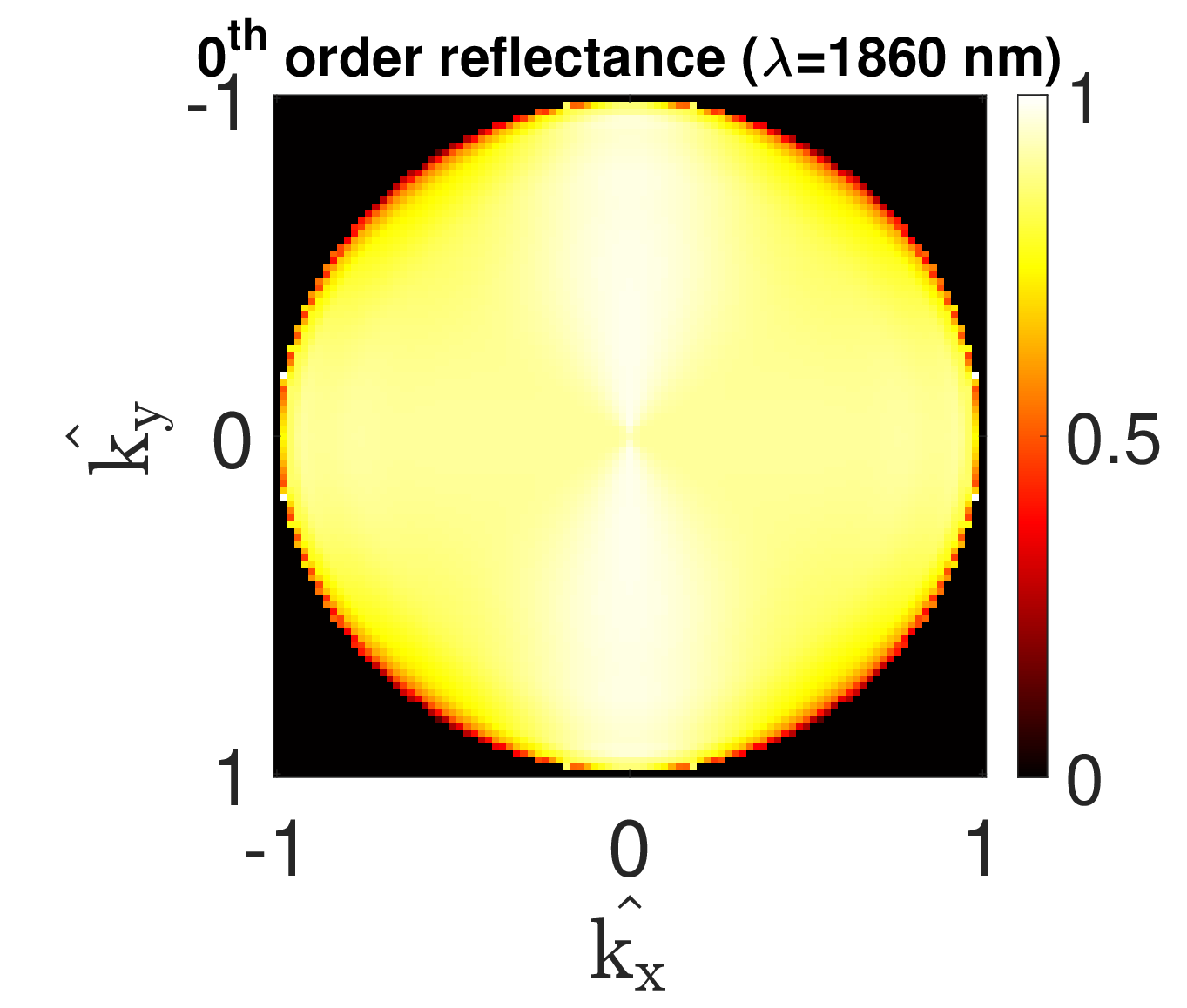}
         \caption{}
         \label{fig:N0_band_exp}
     \end{subfigure}

\centering

     \caption{(a) Angular dispersion of the $0^{\rm th}$-order reflectance, highlighting three wavelengths to demonstrate how the angular invariance range can be controlled by the diffraction cutoff condition. (b) Line plot corresponding to (a), showing strict angular invariance at $\lambda_{0}$ = 1860 nm. (c)–(e) $0^{\rm th}$-order reflectance in full spatial Fourier space for $\lambda_{0}$ = 1400, 1500, and 1860 nm, respectively.}
\label{fig:invariance}
\end{figure}

Beyond the low- and high-pass angular filtering demonstrated earlier, we now show that angular invariance —controllable through diffraction— can be harnessed in reflection to achieve sharp-edge band-pass filtering. Figure \ref{fig:invariance}(a) plots the angular dispersion of the 0$^{\rm th}$-order reflectance for the proposed metagrating with three illustrative wavelengths highlighted: $\lambda_{0}$ = 1.4, 1.5, and 1.86 $\rm \mu m$. The same data are presented in Fig. \ref{fig:invariance}(b) as line graphs, revealing strict angular invariance with respect to $\theta_{\rm inc}$ at $\lambda_{0}$ = 1.86 $\rm \mu m$, with a practical degree of invariance maintained across a broadband spectral range. Figures \ref{fig:invariance}(c)–(e) map the 0$^{\rm th}$-order reflectance over the complete spatial Fourier domain for $\lambda_{0}$ = 1.4, 1.5, and 1.86 $\rm \mu m$, respectively, demonstrating the device’s capability to operate as a reflective Fourier filter. These results confirm that angular invariance in such metagratings enables controllable band-pass filtering with uniform reflectance and sharply defined cutoff angles.

\section{Discussion}
The metagrating results demonstrate how low-pass and high-pass transmissive angular filtering, as well as band-pass reflective angular filtering, can be achieved by exploiting the interplay between Rayleigh anomalies and engineered angular scattering of the unit cell elements, analyzed through the framework of multipolar theory. The physical concept introduced via this metagrating provides a foundation for future studies, including the implementation of arbitrary filtering functions in the spatial Fourier domain.

The potential advantages are significant: it eliminates the need for expensive or impractical lenses, enables unprecedented multiplexing of multiple optical transfer functions across different spectral channels, reduces the footprint by five orders of magnitude, facilitates Fourier filtering of signals with very short coherence lengths, and is compatible with systems involving movable beams without requiring precise alignment. Moreover, the principles demonstrated with this metagrating can be readily extended to other electromagnetic regimes, including microwave, terahertz, and ultraviolet frequencies.

Additionally, the simplicity of the design and its strong tolerance to variations in critical parameters—as detailed in the supplemental materials—underscore the generality of the proposed concept and its broad potential for diverse applications.

\section{Conclusion}
The introduced metagrating establishes the foundational principles for the first compact, planar Fourier optics platform, delivering unprecedented advantages that extend well beyond miniaturization. By enabling the replacement of conventional optical blocks, it opens new possibilities across diverse applications, including imaging, laser systems, and optical computing.

\medskip
\textbf{Supporting Information} \par 


\medskip

%




\section{Disclosure}
The authors declare no conflict of interest.



\pagebreak
\section*{\Huge{Metapinhole: Planar Fourier Optics without lenses: supplemental document}}



\title{Metapinhole: Planar Fourier Optics without lenses: supplemental document}



\section{Angular dispersion of the 0$^{\rm th}$ order transmittance on linear scale}

\begin{figure}[H]
\centering
\begin{subfigure}[b]{0.45\textwidth}
         \centering
         \includegraphics[width=\textwidth]{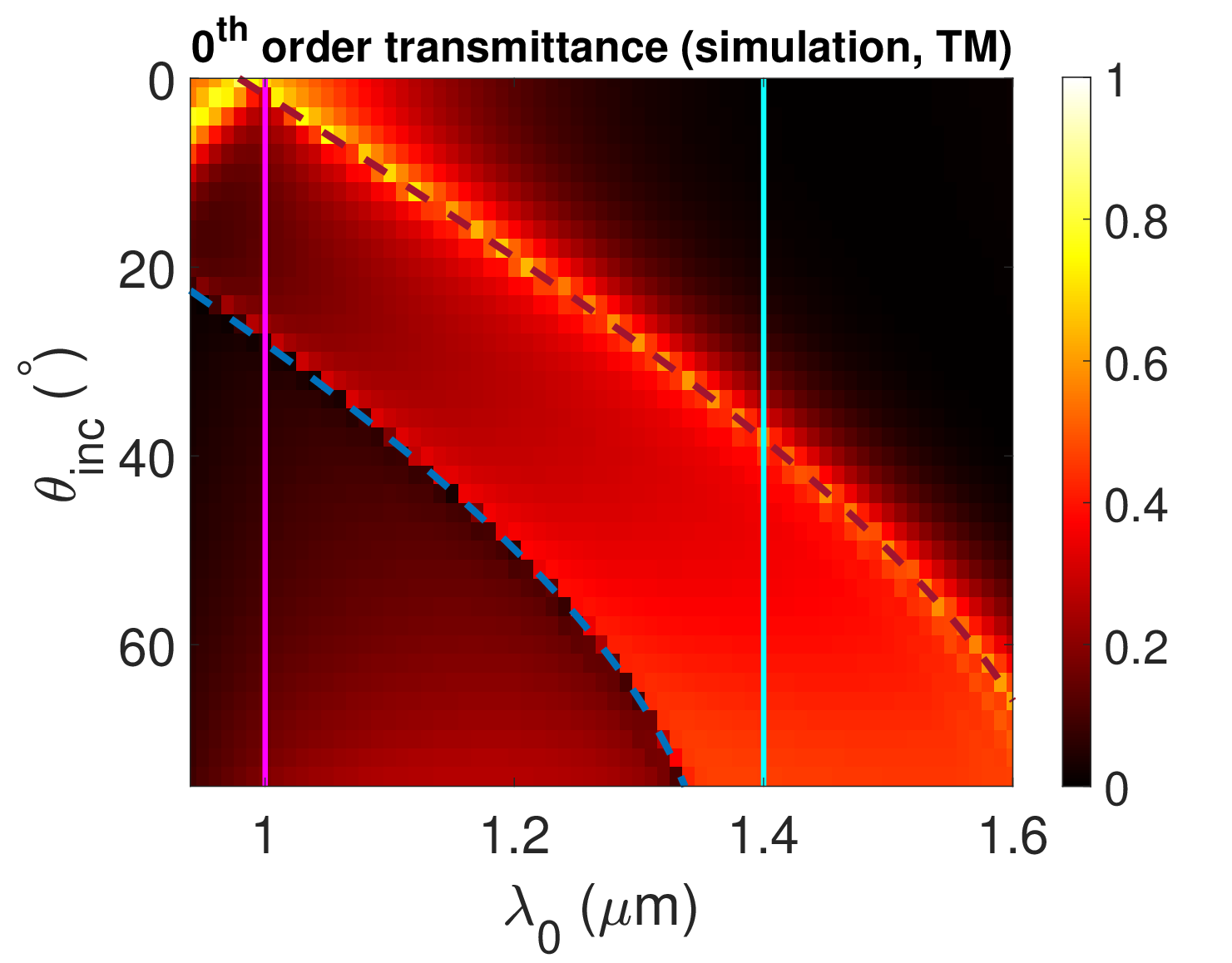}
         \caption{}
         \label{fig:N0_band_theo}
     \end{subfigure}
     \begin{subfigure}[b]{0.45\textwidth}
         \includegraphics[width=\textwidth]{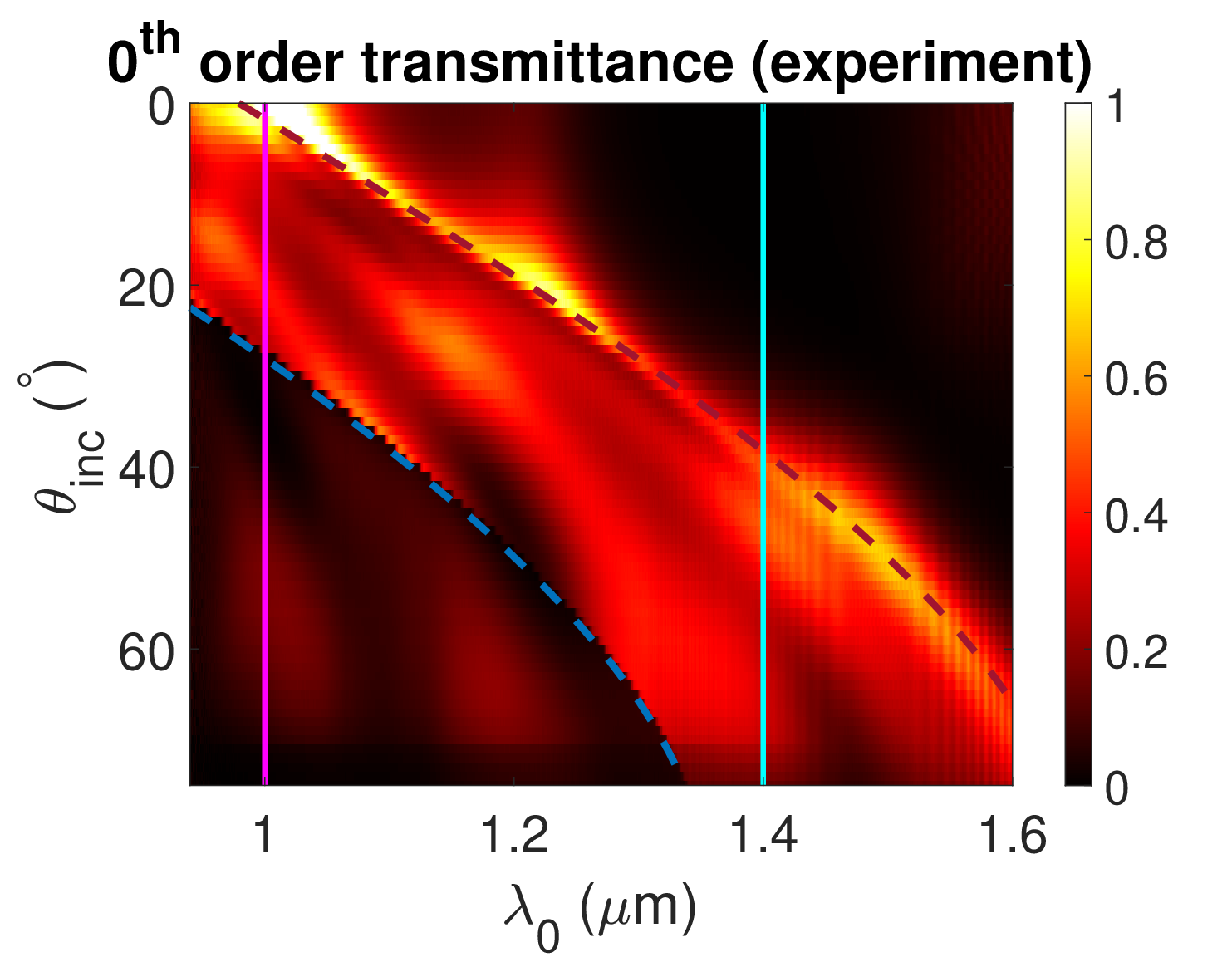}
         \caption{}
         \label{fig:N0_band_exp}
     \end{subfigure} 
     \begin{subfigure}[b]{0.45\textwidth}
         \centering
         \includegraphics[width=\textwidth]{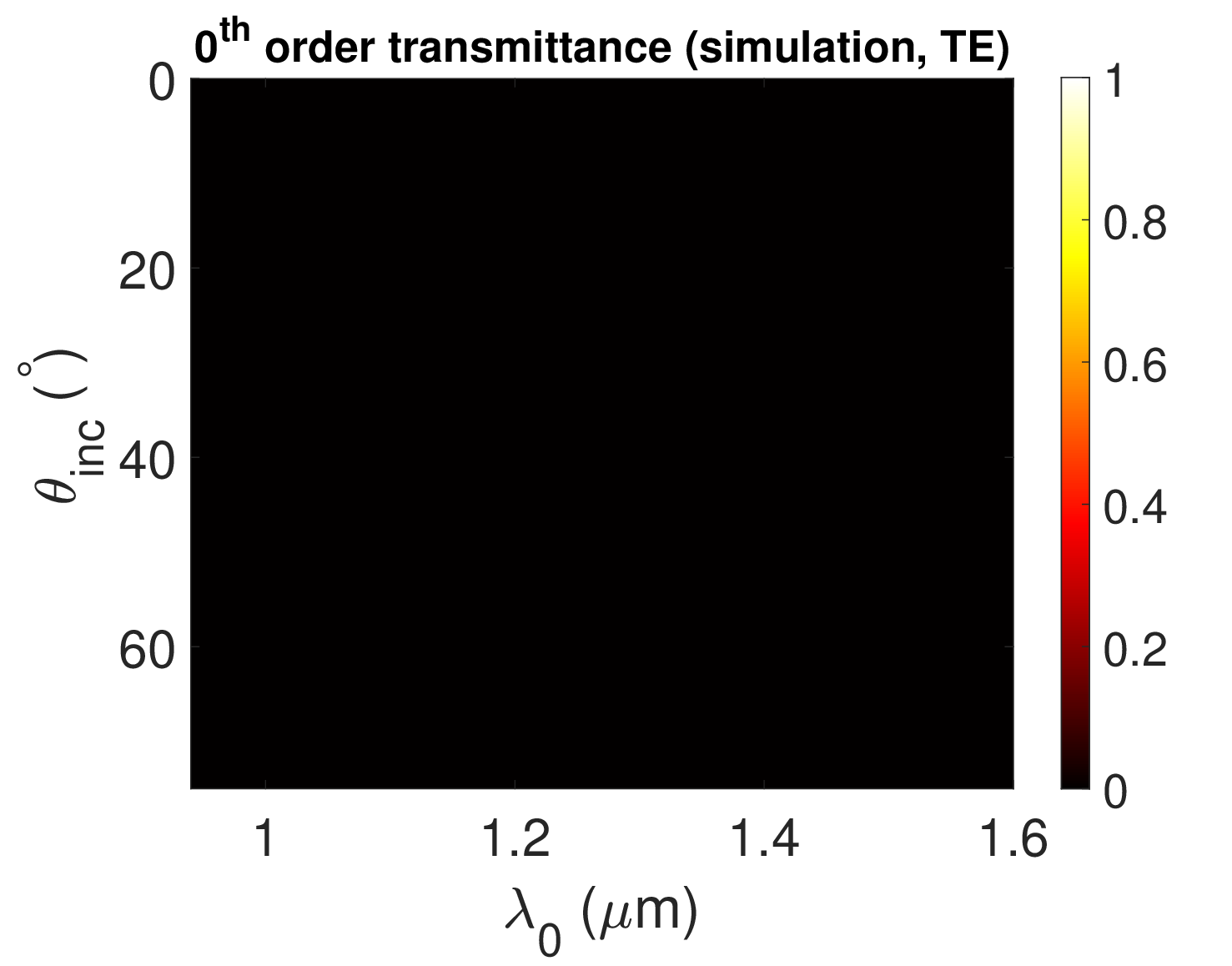}
         \caption{}
         \label{fig:N0_band_theo}
     \end{subfigure}
     \begin{subfigure}[b]{0.45\textwidth}
         \includegraphics[width=\textwidth]{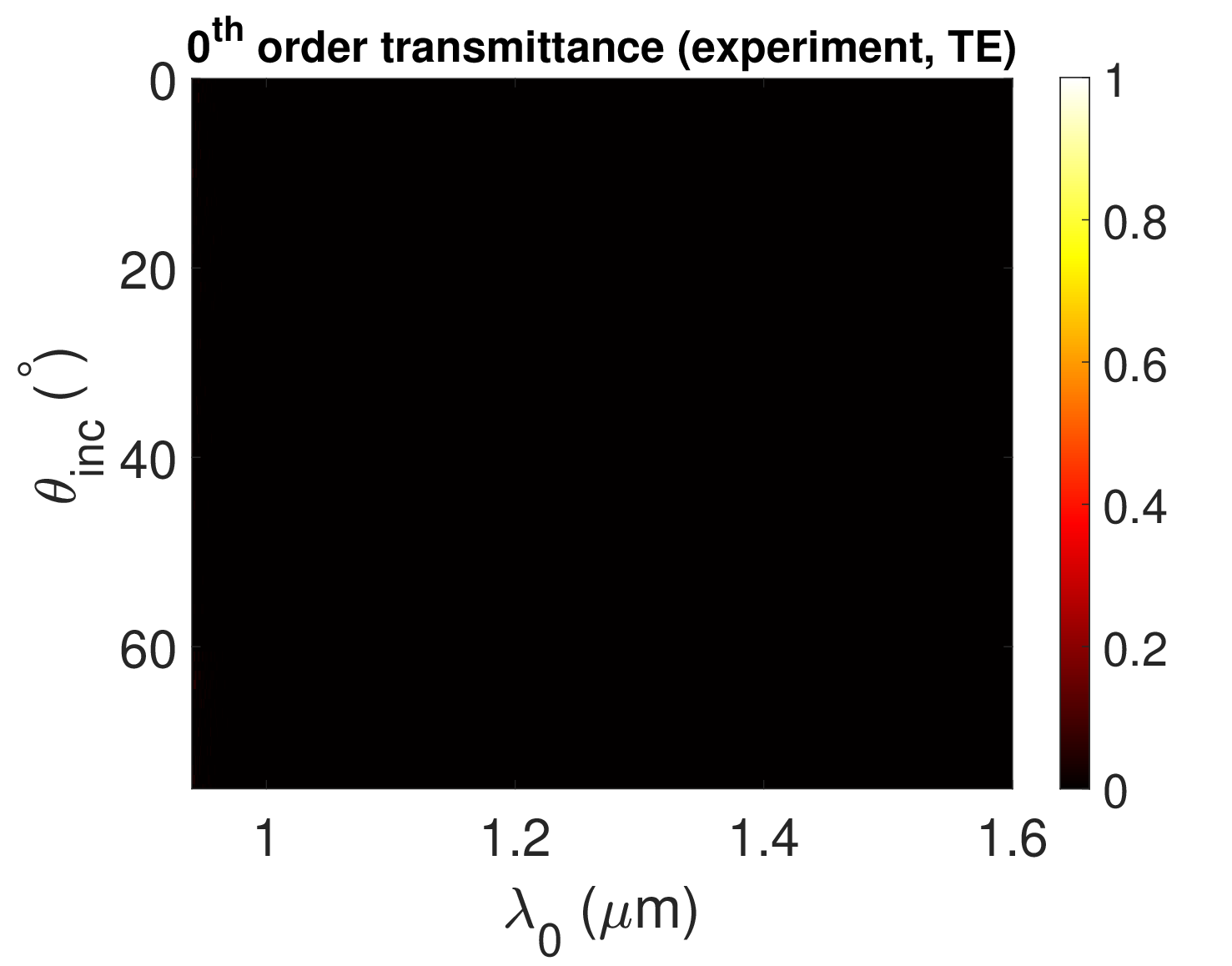}
         \caption{}
         \label{fig:N0_band_exp}
     \end{subfigure} 

\centering

     \caption{(a) Simulated and (b) measured angular dispersion of the $0^{\rm th}$ order transmittance for transverse magnetic (TM) polarization. (c) and (d) Same as (a) and (b) for transverse electric (TE) polarization. The data are identical to Fig.~3 in the main manuscript, but on linear scale. }
\label{fig:linear}
\end{figure}

Figure~\ref{fig:linear} presents the same dataset as Fig.~4 in the main manuscript, but plotted on a linear scale. In this representation, the low-pass filtering effect at $\lambda_{0} = 1 ,\mu\text{m}$ becomes more pronounced, and the strong agreement between simulation and measurement is clearly evident.

\section{ O$^{\rm th}$ order transmittance as a function of spherical angles}
\begin{figure}[H]
\centering
\begin{subfigure}[b]{0.45\textwidth}
         \centering
         \includegraphics[width=\textwidth]{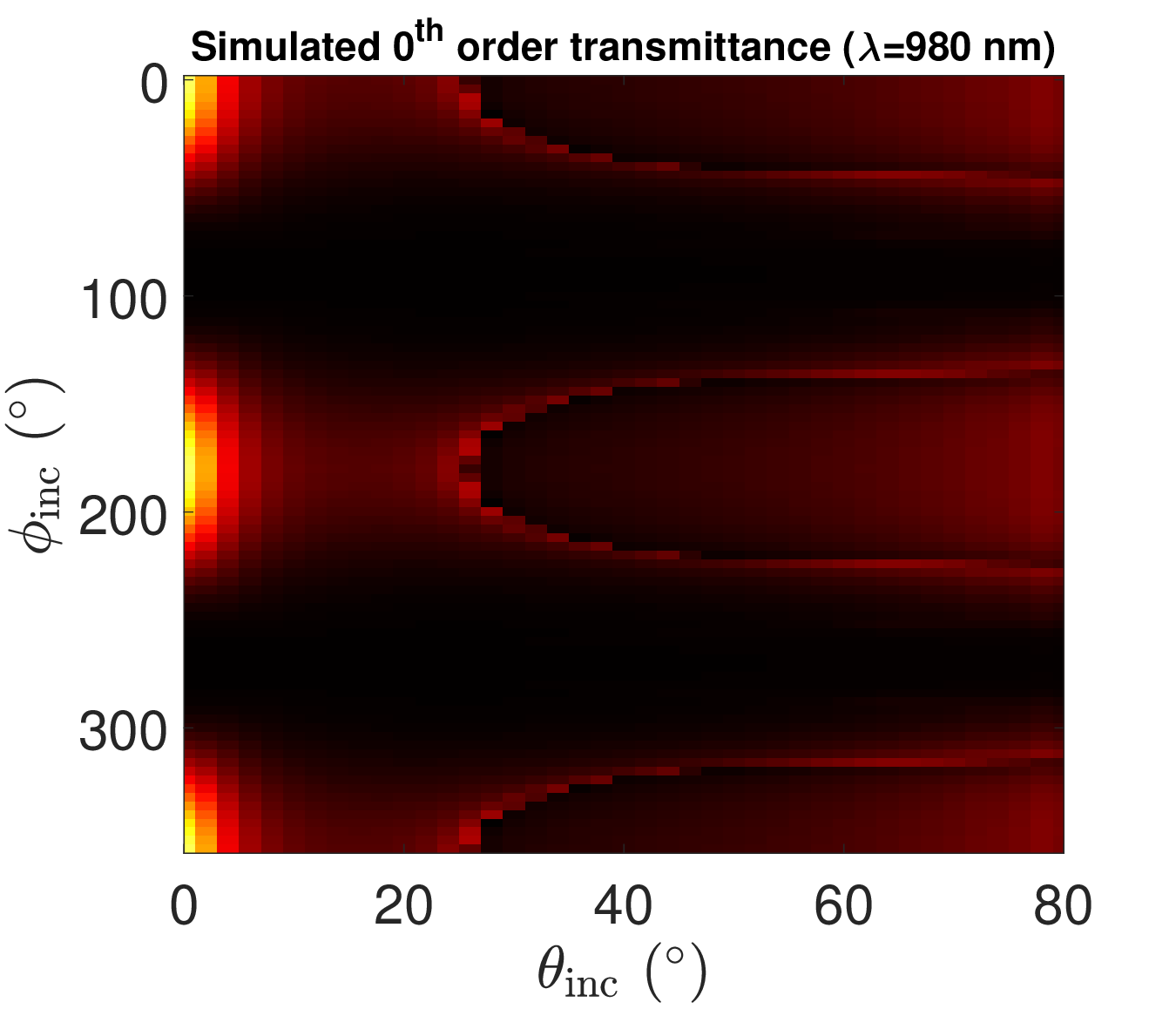}
         \caption{}
         \label{fig:N0_band_theo}
     \end{subfigure}
     \begin{subfigure}[b]{0.45\textwidth}
         \includegraphics[width=\textwidth]{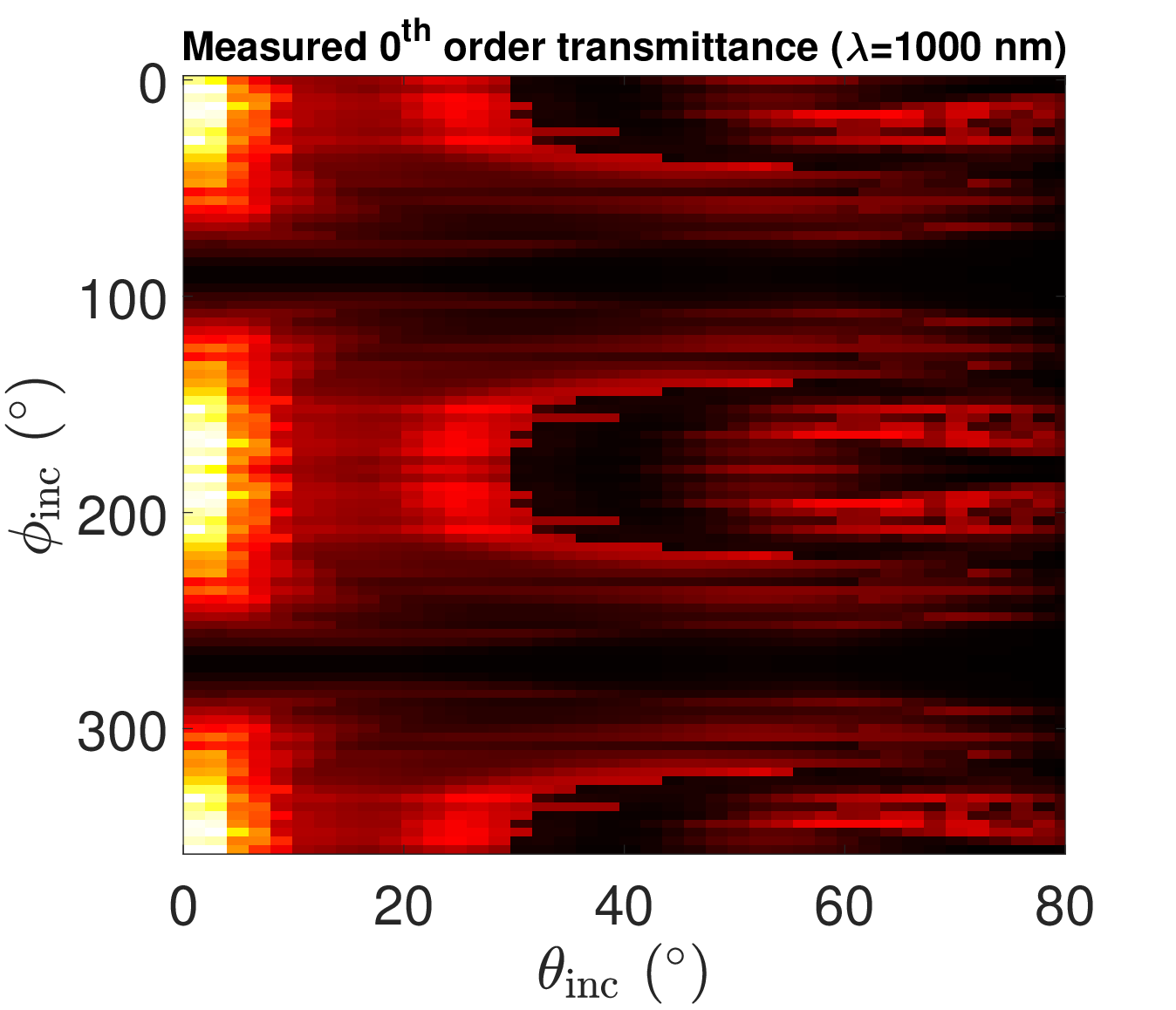}
         \caption{}
         \label{fig:N0_band_exp}
     \end{subfigure} 
     \begin{subfigure}[b]{0.45\textwidth}
         \centering
         \includegraphics[width=\textwidth]{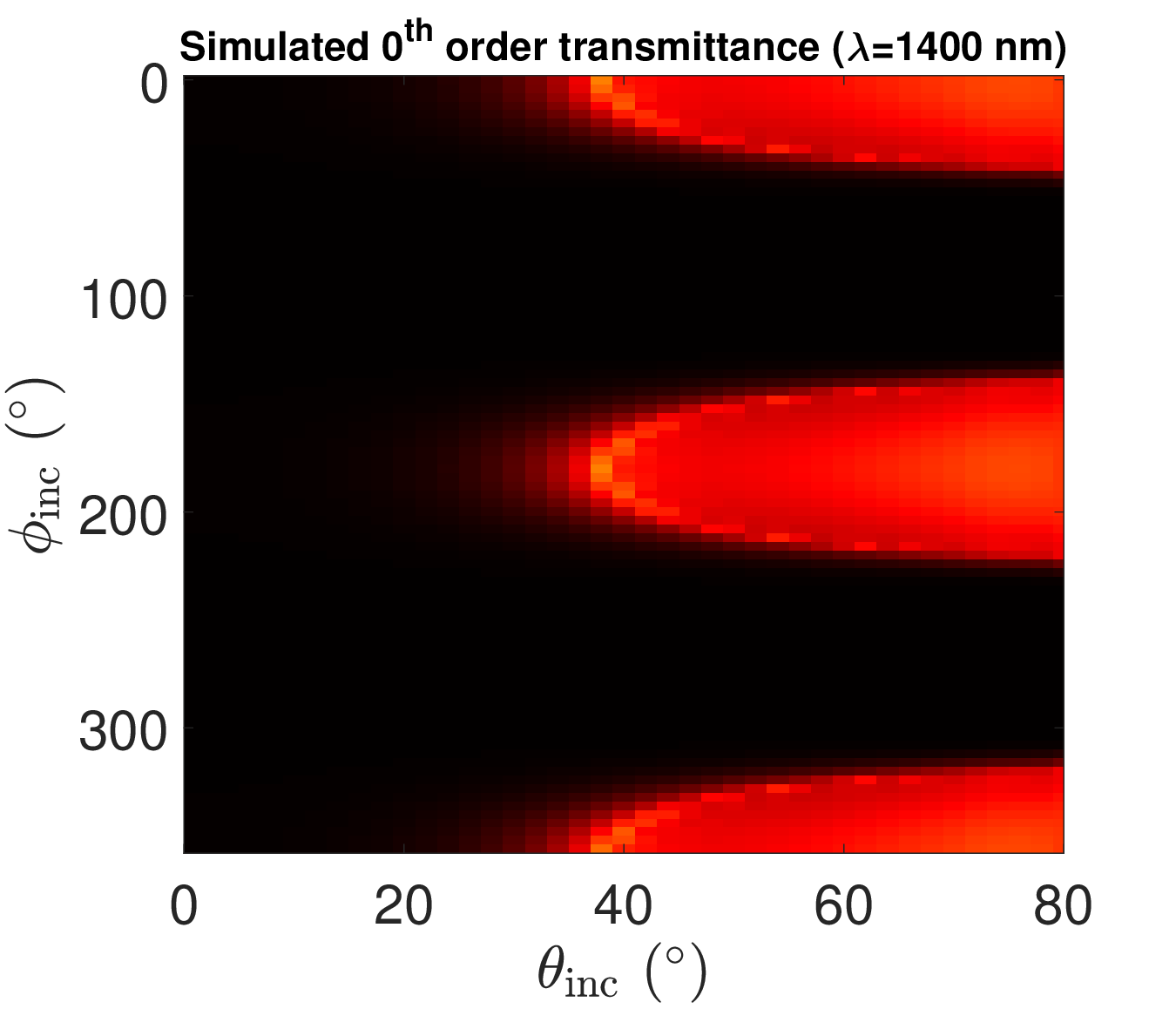}
         \caption{}
         \label{fig:N0_band_theo}
     \end{subfigure}
     \begin{subfigure}[b]{0.45\textwidth}
         \includegraphics[width=\textwidth]{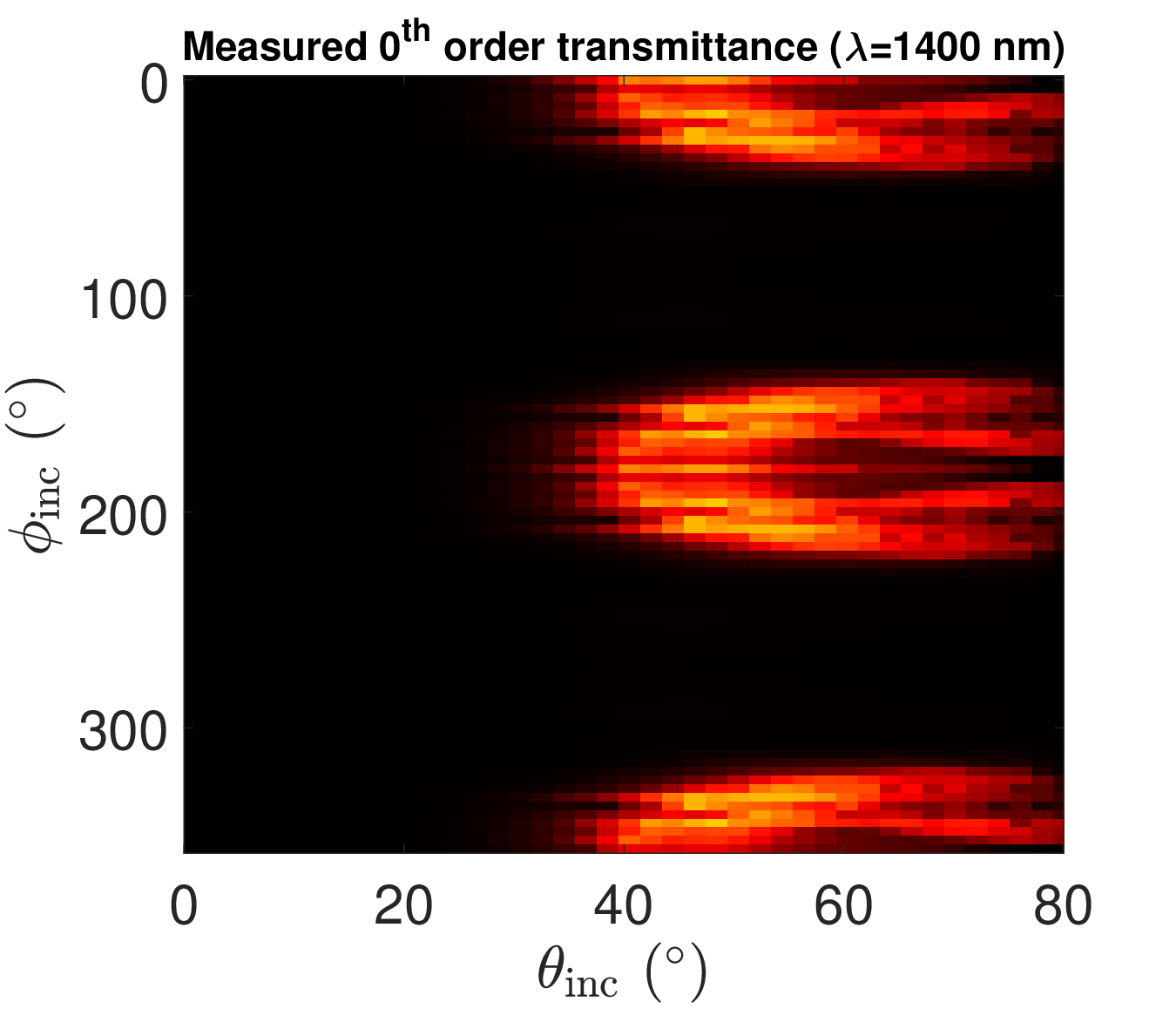}
         \caption{}
         \label{fig:N0_band_exp}
     \end{subfigure} 

\centering

     \caption{(a) Simulated and (b) measured $0^{\rm th}$-order transmittance for a low-pass angular filter  as a function of all three-dimensional angles at $\lambda_{0}$ = 980 nm and 1000 nm, with a 20 nm shift between simulation and measurement. (c) and (d) Same as (a) and (b), but for a high-pass angular filter at $\lambda_{0}$ = 1400 nm. }
\label{fig:spherical}
\end{figure}

Figure~\ref{fig:spherical} presents the original simulation and measurement data, resolved point-by-point for all three-dimensional spherical angles of incidence, in order to compute the metagrating’s response across the full spatial Fourier space via spherical-to-Cartesian transformation (i.e., Fig.~5 in the main manuscript). This figure is based on 1,890 individual measurements, with all data normalized to the transmittance of a double-sided polished silicon wafer at normal incidence. The measured dataset covers only the angular range $\phi_{\mathrm{inc}} = 0^\circ$ to $90^\circ$, while the remaining values for $\phi_{\mathrm{inc}} = 90^\circ$ to $360^\circ$ are reconstructed by exploiting the spatial symmetries of the metagrating.

\section{Fabrication and measurements}
\begin{figure}
\centering
\includegraphics[width=0.8\textwidth]{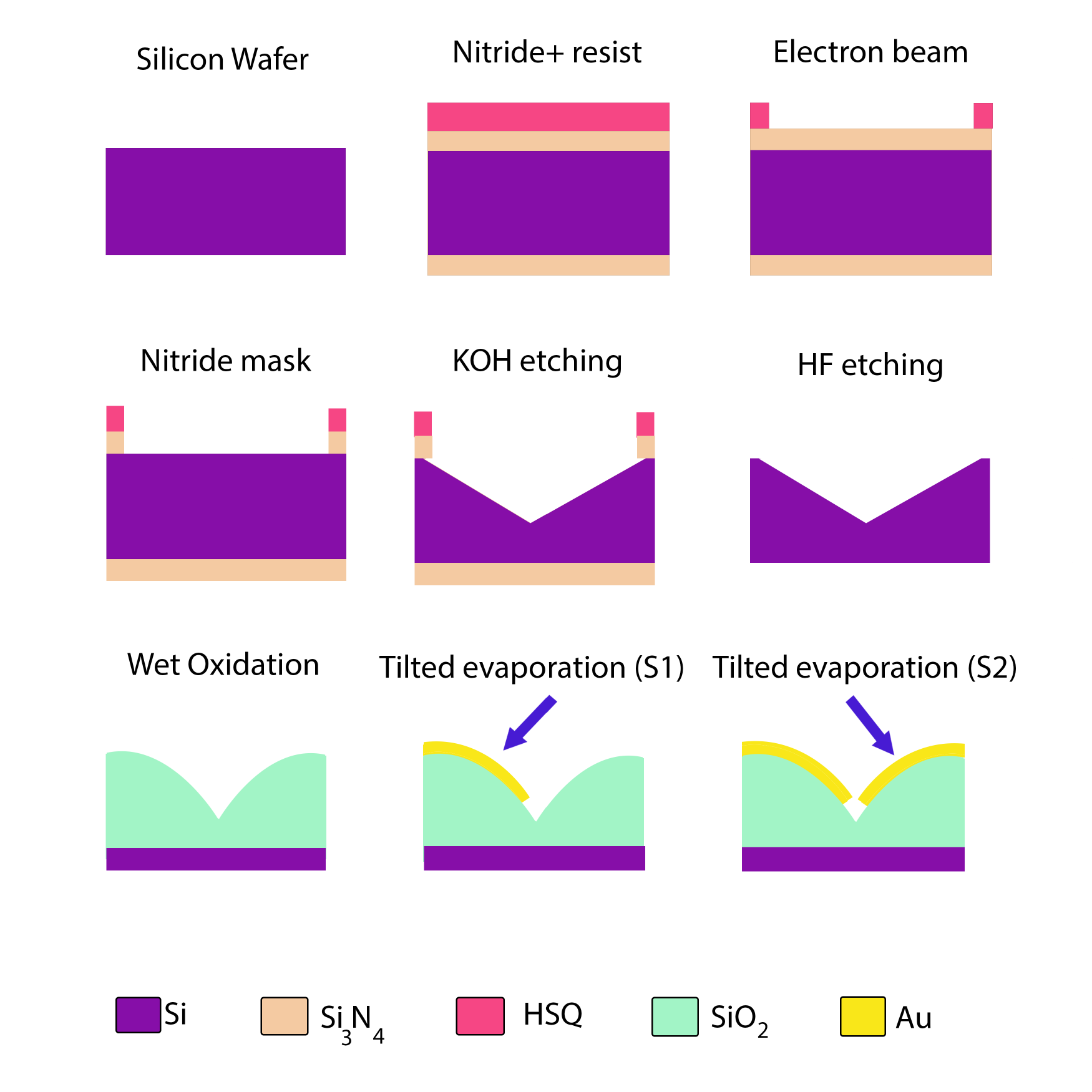}
\caption{Process flow for the fabrication.}
\label{fig:process_flow}
\end{figure}

Figure~\ref{fig:process_flow} illustrates the fabrication process of the metagrating. First, a 10-nm-thick silicon nitride film is deposited on a silicon wafer using low-pressure chemical vapor deposition (LPCVD). Electron-beam lithography followed by plasma etching is employed to define the nitride mask. The sample is then immersed in a 40\% KOH solution to form an array of triangular ridges~\cite{step}. Subsequent wet thermal oxidation converts these ridges into a reversed parabolic silica pattern. Finally, two steps of oblique-angle evaporation, at $70^\circ$ and $-70^\circ$, are performed to deposit a hat-like gold layer with well-defined gaps on top of the silica ridges.

\begin{figure}
\centering
\includegraphics[width=\textwidth]{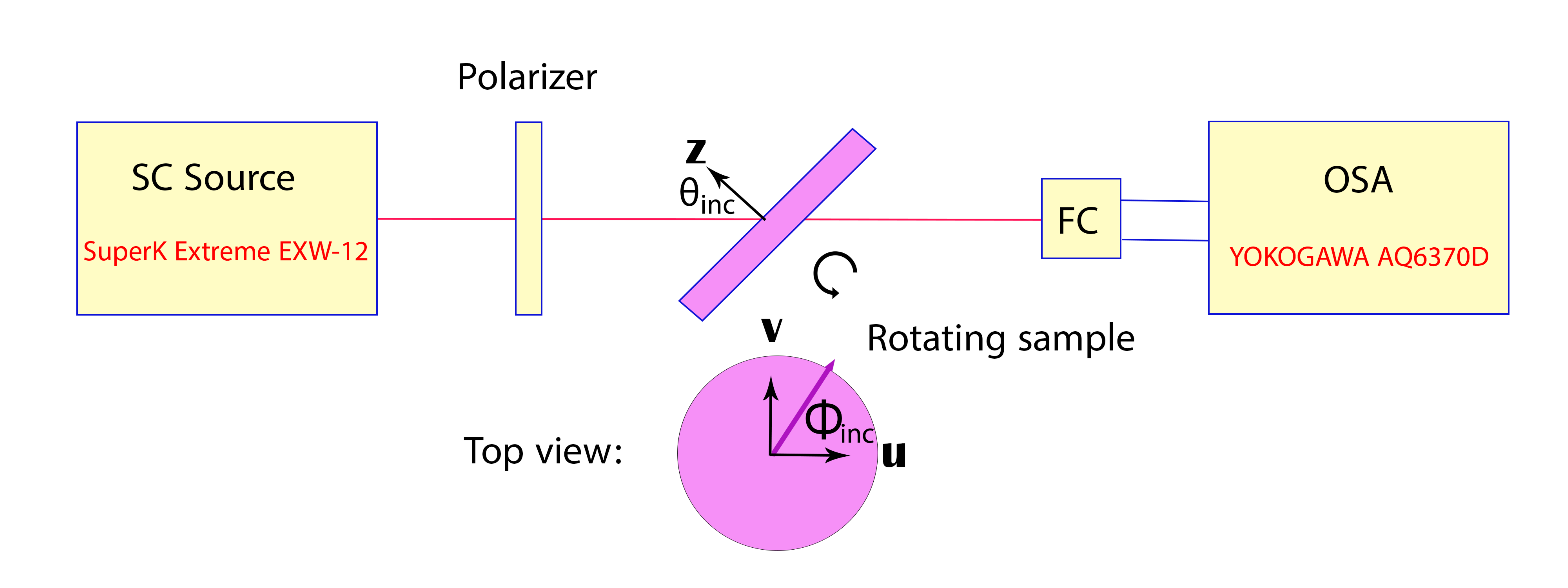}
\caption{Optical Setup for measuring 0$^{\rm th}$ order transmittance.}
\label{fig:setup}
\end{figure}

Figure~\ref{fig:setup} shows the optical setup used for the measurements. A supercontinuum source (Super K Extreme EXW-12, NKT Photonics) provides the illumination. A polarizer, mounted on a rotating holder capable of accessing all spherical angles ($\theta_{\mathrm{inc}}$ and $\phi_{\mathrm{inc}}$), is used to control both the incident polarization and the orientation of the metagrating. An optical spectrum analyzer (YOKOGAWA AQ6370D), connected via a fiber coupler (FC), records the spectral response across all wavelengths in a single measurement. The violet arrow in the top view indicates the in-plane axis perpendicular to the hat-like gold cylinders (corresponding to the $x$-axis in Fig.~2 of the main manuscript). The $z$-axis is always defined as perpendicular to the surface of the metagrating.

\section{Insensitivity to geometrical variations}

\begin{figure}
\centering
\begin{subfigure}[b]{0.45\textwidth}
         \centering
         \includegraphics[width=\textwidth]{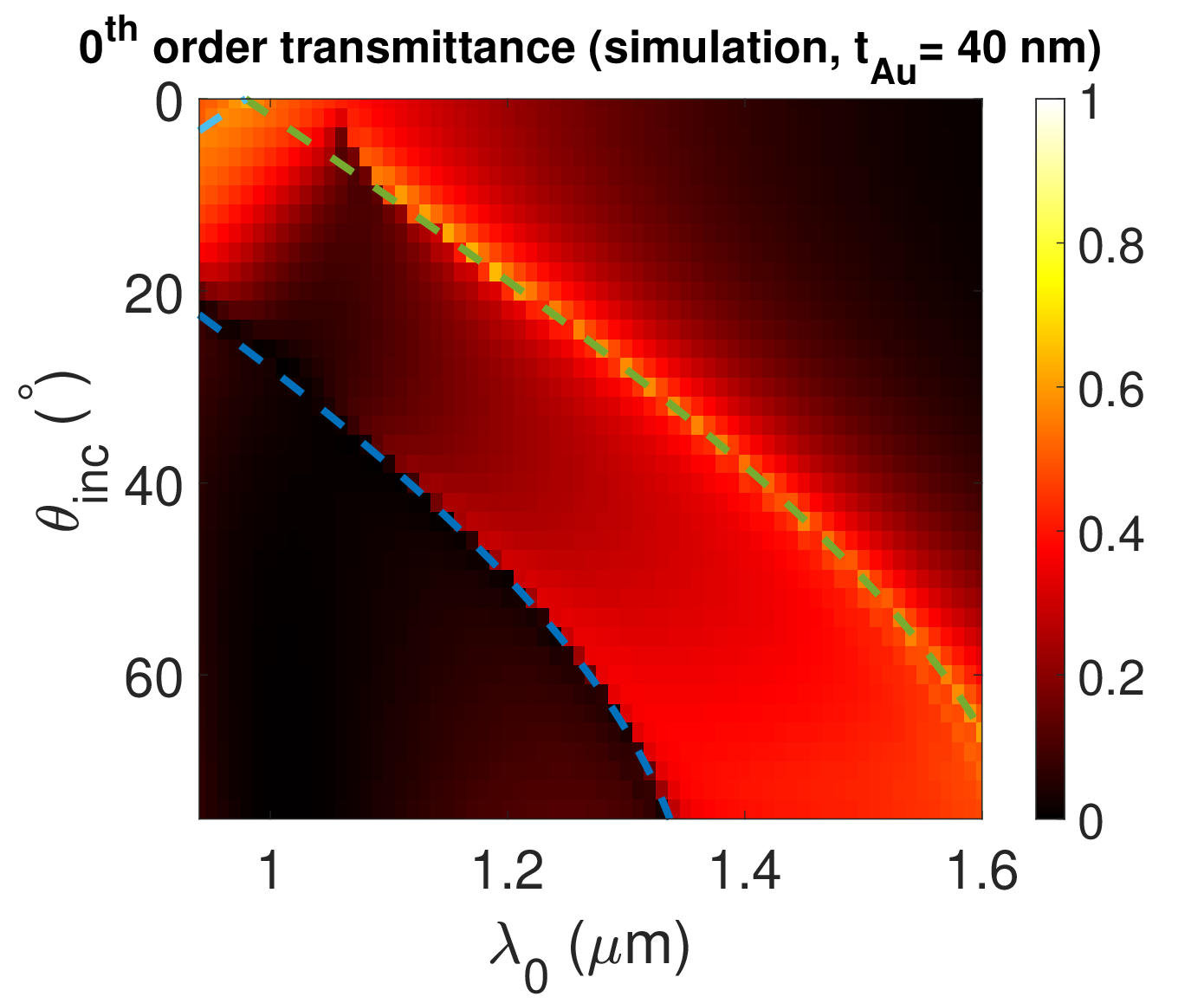}
         \caption{}
         \label{fig:N0_band_theo}
     \end{subfigure}
     \begin{subfigure}[b]{0.45\textwidth}
         \includegraphics[width=\textwidth]{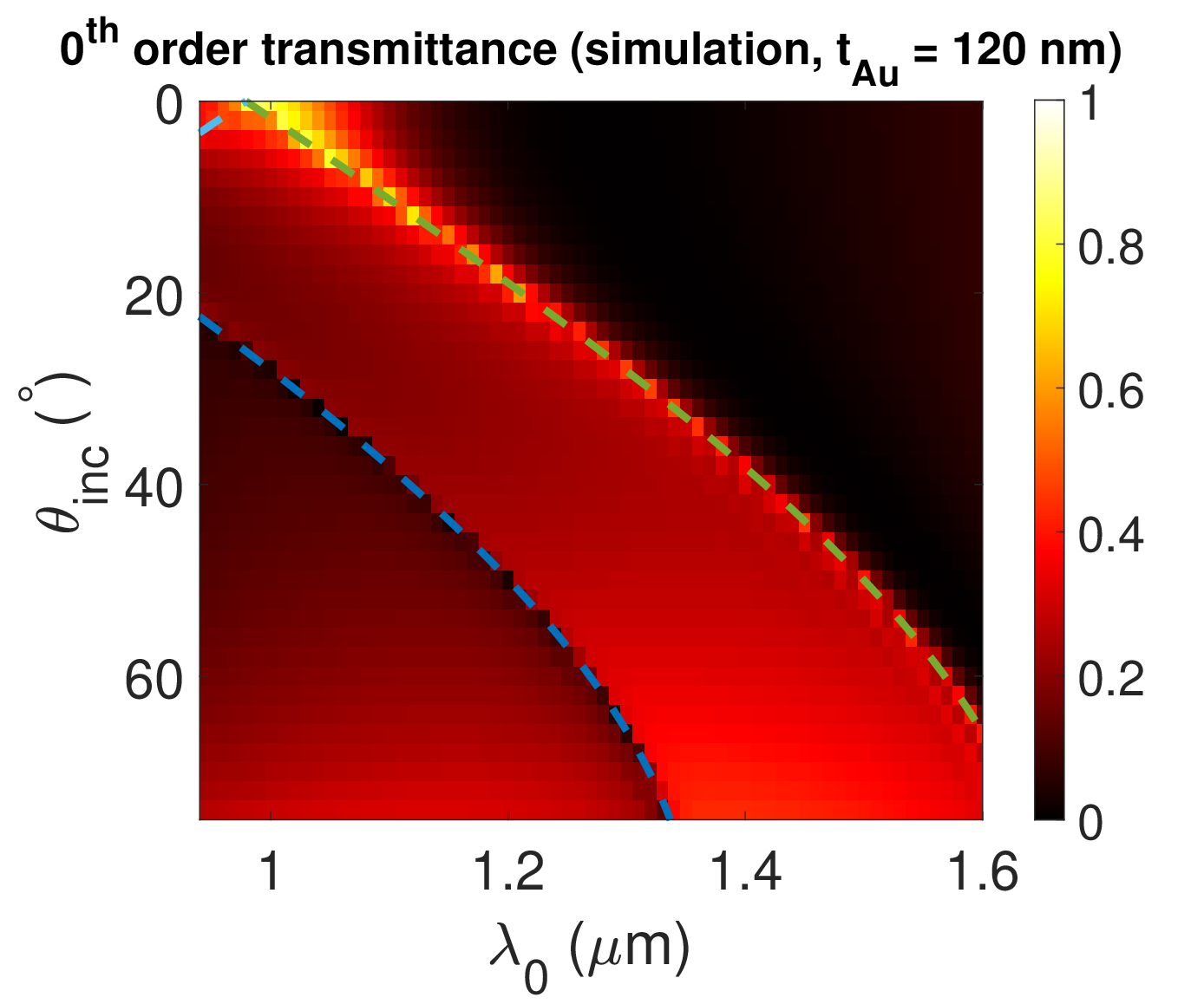}
         \caption{}
         \label{fig:N0_band_exp}
     \end{subfigure} 
     \begin{subfigure}[b]{0.45\textwidth}
         \centering
         \includegraphics[width=\textwidth]{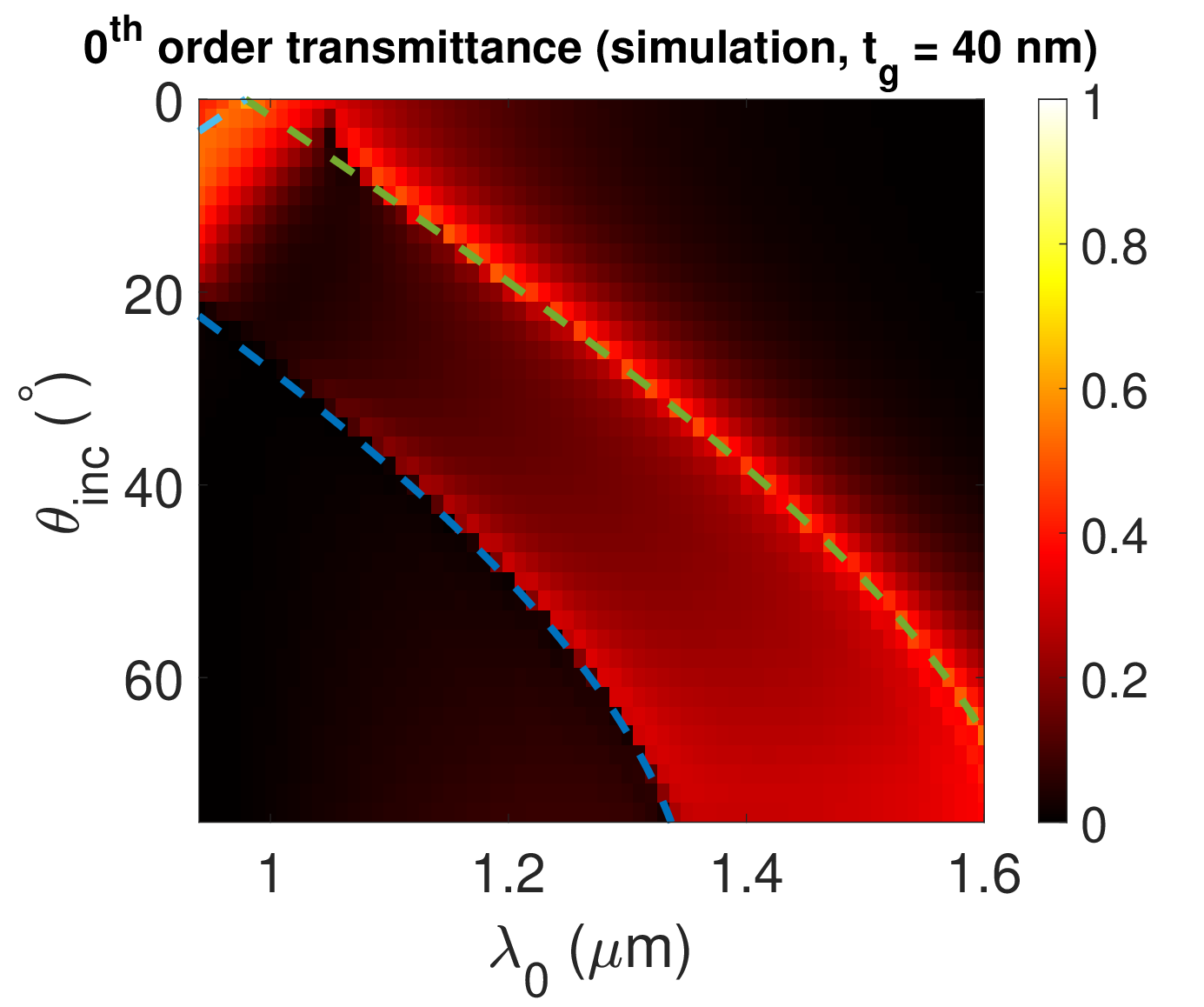}
         \caption{}
         \label{fig:N0_band_theo}
     \end{subfigure}
     \begin{subfigure}[b]{0.45\textwidth}
         \includegraphics[width=\textwidth]{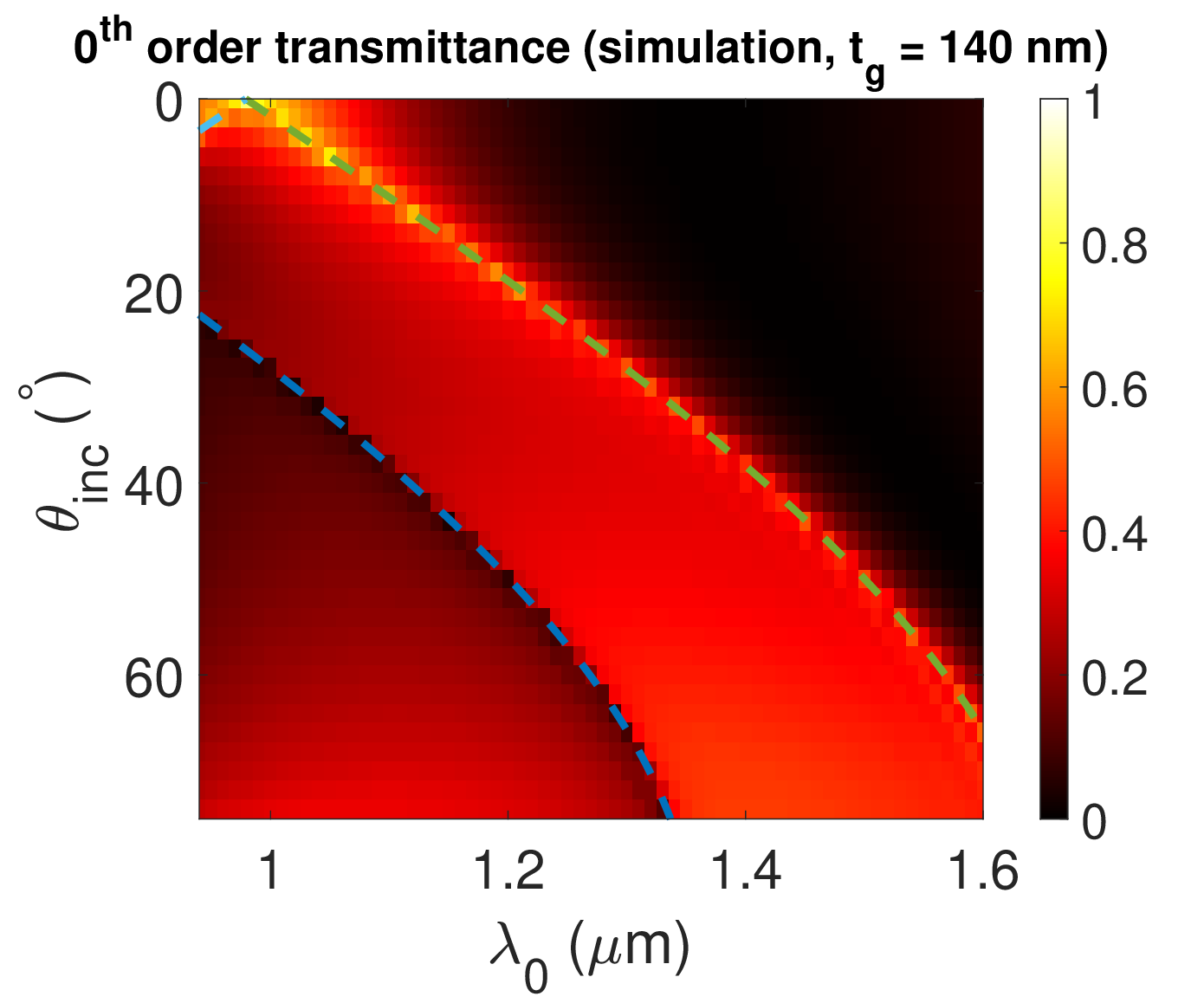}
         \caption{}
         \label{fig:N0_band_exp}
     \end{subfigure} 

\centering

     \caption{$0^{\rm th}$-order transmittance of the metagrating with the same parameters as Fig.~2 in the main manuscript, except with one extreme geometrical variation: (a) gold thickness $t_{Au} = 40$ nm, (b) $t_{Au} = 120$ nm, (c) gap width between two hat-like elements $t_{\rm g} = 40$ nm, and (d) $t_{\rm g} = 140$ nm.}
\label{fig:insens}
\end{figure}

Figure~\ref{fig:insens} presents the angular dispersion of the metagrating introduced in Fig.2 of the main manuscript under TM polarization, subject to extreme deviations in a single critical geometrical parameter. For comparison with the reference gold thickness ($t_{\rm Au} = 78$ nm), Figs.\ref{fig:insens}(a) and (b) show the results for $t_{\rm Au} = 40$ nm and $t_{\rm Au} = 120$ nm, respectively. Similarly, relative to the reference gap width between adjacent hat-like elements ($t_{\rm g} = 80$ nm), Figs.~\ref{fig:insens} (c) and (d) illustrate the cases of $t_{\rm g} = 40$ nm and $t_{\rm g} = 140$ nm. Remarkably, the angular dispersion remains largely unaffected despite such extreme variations, highlighting the resilience of the proposed metagrating against wafer-scale fabrication imperfections. This robustness underscores the practicality of the structure, whose optical response originates from its topology rather than from fine-tuned geometrical parameters.



\end{document}